\documentclass[prb,aps,final,twocolumn,floatfix]{revtex4}
\input epsf
\def\diagp#1#2{\vbox to 1.2cm{
\hbox to #2cm{
\hfill
\includegraphics{#1.eps}}
\vfill}}

\def\diag#1#2{\vbox to 1cm{
\hbox to #2cm{
\hfill
\includegraphics{#1.eps}}
\vfill}}
\def\diagq#1#2{\vbox to 1cm{
\hbox to #2cm{
\hfill
\includegraphics{#1.eps}}
\vfill}}

\def\diagg#1#2#3{\raisebox{-#3ex}{\setlength{\epsfxsize}{#2cm}
\leavevmode \epsfbox{#1.eps}}}

\begin{document}
\title{Theoretical study of finite temperature spectroscopy in van der
Waals clusters. I. Probing phase changes in CaAr$_n$}
\author{F. Calvo, F. Spiegelman and M.-C. Heitz}
\affiliation{Laboratoire de Physique Quantique, IRSAMC, Universit\'e Paul
Sabatier, 118 Route de Narbonne, F31062 Toulouse Cedex, France}
\begin{abstract}
The photoabsorption spectra of calcium-doped argon clusters CaAr$_n$ are
investigated at thermal equilibrium using a variety of theoretical and
numerical
tools. The influence of temperature on the absorption spectra is estimated
using the quantum superposition method for a variety of cluster sizes in the
range $6\leq n\leq 146$.
At the harmonic level of approximation, the absorption intensity is
calculated through an extension of the Gaussian theory by Wadi and Pollak
[J. Chem. Phys. {\bf 110}, 11890 (1999)]. This theory is tested on simple,
few-atom systems in both the classical and quantum regimes for which highly
accurate Monte Carlo data can be obtained. By incorporating quantum
anharmonic corrections to the partition functions and respective weights of
the isomers, we show that
the superposition method can correctly describe the finite-temperature
spectroscopic properties of CaAr$_n$ systems. The use of the absorption
spectrum as a possible probe of isomerization or phase changes in the argon
cluster is discussed at the light of finite-size effects.
\end{abstract}
\maketitle

\section{Introduction}

Large van der Waals clusters offer a convenient medium to study solvation at
the molecular level, and also to undertake chemical reactions with well
defined thermodynamical conditions.\cite{cicr1,cicr2,cicr3} Reactivity of
chromophore-doped rare-gas clusters has been the subject of intense
experimental effort.%
\cite{baumfalk00,slavicek,baumfalk01,gaveau00,liu01,briant01}
Photodissociation of HBr on argon clusters has been investigated by Baumfalk
{\em et al.},\cite{baumfalk00,slavicek,baumfalk01} including a theoretical
study of the HBr--cluster interaction.\cite{slavicek} Chemiluminescence in the
bimolecular reaction Ba$+$N$_2$O$\longrightarrow$BaO$^*+$N$_2$ was the subject
of a work by Gaveau {\em et al.},\cite{gaveau00} and photodetachment in
IHI$^-$Ar$_n$ clusters was used to probe transition state spectroscopy of the
I$+$HI reaction by Liu and coworkers.\cite{liu01} One may also cite the
photoinduced reaction
dynamics between Ca$^*$ and HBr towards CaBr$^*+$H at the surface of a large
argon cluster carried out by Briant {\em et al.}\cite{briant01} Very
recently, the reaction between N$_2$O and Li$_2$ mediated by Ar$_n$ clusters
has been investigated by Breckenridge and coworkers.\cite{breck02}

A practical understanding of the physical and chemical properties of the
deposited species relies on a good characterization of their spectroscopy.
In turn, spectroscopy gives great insight into the structure of the van der
Waals complex, especially in concern with the location of the chromophore.%
\cite{depujo92,depujo94,epjd}
This idea was used to propose a way of detecting structural transformations
inside the cluster.\cite{hahn,even} Not long after its was discovered in
simulations,\cite{berry} the so-called dynamical coexistence phenomenon
between solidlike and liquidlike phases
received some experimental evidence by Hahn and Whetten who reported changes
in the excitation spectrum of benzene-(argon)$_n$ clusters with increasing
temperature.\cite{hahn} Similar
experimental results were reported for dichloroanthracene-Ar$_n$ clusters by
Even {\em et al.}\cite{even} More recently, Curotto and
coworkers\cite{curotto1,curotto2,curotto3} theoretically investigated the
infrared spectroscopy in Ar$_n$-HF clusters. They observed significant
variations in the red shift associated with the two lowest vibrational states
of the HF molecule, which they could correlate with the thermodynamical state
of the cluster.\cite{curotto1,curotto2,curotto3} Moseler {\em et
al.}\cite{moseler} found that the
photoabsorption spectra of Na$_8$ displays some features characteristic of
isomerization and structural fluctuations. Lastly, simulations on charged
rare-gas clusters have investigated the influence of temperature on the
absorption spectrum.\cite{grigorov,galindez} So far, no direct evidence for
isomerization in such rare-gas clusters has been found yet.

While spectroscopy has many attractive features for experimental studies in
large clusters, several difficulties pave the way for efficient theoretical
approaches. Firstly, because of the numerous degrees of freedom, a fully
quantum calculation of the absorption spectrum by exact methods (wavepacket
or diagonalization in a basis)
is not feasible beyond a few atoms and one is usually lead to simplify the
problem. A common approximation is the Condon hypothesis, which
assumes vertical transitions. Secondly, the equilibrium properties at finite
temperature require some sampling of the ground state
potential energy surface (PES). This
sampling can be hard to get because of the topography of the multidimensional
PES (the so-called energy landscape). When the energy landscape is complicated
or has multiple funnels, the relaxation to equilibrium may be particularly
slow with a glassy-like behavior.\cite{walesacp} Broken
ergodicity is then likely to occur in conventional molecular dynamics (MD) or
Monte Carlo (MC) simulations, and also in quantum simulations.
Fortunately some remedies to this problem have recently been suggested,
including multicanonical ensemble sampling\cite{berg} or parallel
tempering.\cite{ptmc}
Thirdly, most experiments involving chromophore-doped rare-gas
clusters take place at low temperatures, at which the classical approximation
may be questioned, especially for the lighter elements neon and helium.
Path-integral Monte Carlo or centroid MD should then
be used instead of classical
techniques, however their use with ergodicity restoring strategies remains
numerically demanding for systems containing tens or hundreds degrees of
freedom.

We have recently proposed an alternative way of calculating thermodynamical
observables at canonical equilibrium, using the superposition approximation of
quantum oscillators.\cite{jcpquantum} This method, originally developed for
thermodynamical observables only, was improved phenomenologically by including
anharmonic corrections,\cite{jcpquantum,anhardoye} and extended to generally
treat other temperature-dependent
observables.\cite{prop} Its main interest is to circumvent the troubles of
broken ergodicity by considering separately the contribution of each isomer
to the partition function. The method also naturally describes quantum
delocalization in a straightforward way.

One goal of this paper is to extend the quantum superposition method to the
realm of calculating photoabsorption spectra. In addition to the ingredients
presently available, one needs to calculate the photoabsorption spectrua
of each stable isomer. For this, we use a recent theory by Wadi and
Pollak\cite{wp} that we have extended to the general case of arbitrary
potential surfaces. As will be seen below, this Gaussian theory of absorption
succeeds in predicting absorption line shapes, widths and shifts for harmonic
polyatomic molecules in good agreement with highly accurate Monte Carlo data.

With the quantum superposition method, we are able to carry out
systematic studies for
a given set of cluster sizes. We can then address our second goal, and try to
relate the features of the absorption spectrum to the knowledge of
isomerizations and possible phase changes inside the cluster.
More specifically,
by looking at particular sizes we hope to gain insight in the relationship
between structure, temperature, quantum delocalization, and the shape and
position of the spectral lines.

This work follows previous efforts aimed at interpreting experimental
absorption spectra on large CaAr$_n$ clusters.\cite{epjd} The spectroscopy
of the CaAr diatomic molecule has been investigated both experimentally and
with {\em ab initio} calculations in Ref.~\onlinecite{abinit}, and a
Diatomic-In-Molecules (DIM) Hamiltonian was constructed to model the $4s$
ground- and $4s4p$ excited-states of CaAr$_n$ clusters.\cite{epjd} In the
present context of finite-temperature spectroscopy, the heavy numerical cost of
statistical sampling is balanced with this relatively cheap, albeit accurate
Hamiltonian. The paper is organized as follows. In the next section, we briefly
recall the quantum superposition method in the harmonic approximation. We also
include the derivation of approximate perturbative anharmonic quantum
corrections to the partition functions. The Gaussian theory of absorption for
harmonic molecules,
initially proposed by Wadi and Pollak,\cite{wp} will be generalized to include
possible Duschinskii rotations in Sec.~\ref{sec:gaussian}. These methods will
be tested on simple clusters, in both the classical and quantum regimes, by
comparing their results to high statistics Monte Carlo
data. Their main application to the variations of the photoabsorption spectrum
with temperature in some size-specific CaAr$_n$ clusters will be presented in
Sec.~\ref{sec:appl}. The results will be discussed in view of the isomerization
and phase changes phenomena. Some concluding remarks will close the paper in
Sec.~\ref{sec:ccl}.

\section{Quantum superposition method}
\label{sec:super}

In the superposition approach,\cite{hoare79,stillinger82,wales93} also known
as the
multiple normal modes model,\cite{franke} the total partition function $Z$
of the $N-$atom system in equilibrium at temperature $T=1/k_B\beta$ is replaced
by a sum over stable minima $\{\alpha\}$:
\begin{equation}
Z(\beta)=\sum_\alpha n_\alpha Z_\alpha(\beta).
\label{eq:zinit}
\end{equation}
Each minimum $\alpha$, or inherent structure, has a contribution $Z_\alpha$
to $Z$, further weighted by the degeneracy factor $n_\alpha$, which accounts
for permutational isomers and global symmetry. In CaAr$_n$ clusters ($N=n+1$),
$n_\alpha$ is given by $2n!/h_\alpha$, where
$h_\alpha$ is the order of the point group of structure $\alpha$. The problem
has now been reduced to estimating the individual functions $Z_\alpha$. At the
simplest level of approximation, the harmonic vibrational frequencies $\{
\omega_{\alpha i}\}$ provide a zero-th order expression for $Z_\alpha$:
\begin{equation}
Z_\alpha^{(0)}(\beta) = e^{-\beta E_\alpha} \prod_i \frac{2}{{\rm sinh}\, \beta
\hbar \omega_{\alpha i}/2},
\label{eq:z0}
\end{equation}
with $\hbar$ the reduced Planck's constant and $E_\alpha$ the potential energy
minimum of isomer $\alpha$. In the classical regime $\hbar\to 0$, the
decomposition (\ref{eq:zinit}) is formally exact, since each point of the
configuration space (except saddle points) belong to one and only one
basin $\alpha$. In the quantum regime (finite $\hbar$), tunnelling is
neglected.\cite{jcpquantum} In principle, the knowledge of all minima of the
energy landscape allows us to calculate any thermodynamical observable,
including internal energy and heat capacity, from Eq.~(\ref{eq:zinit}).
However, for all but the smallest atomic systems, exhaustive enumeration of
these isomers is not possible because the number of local minima is thought to
grow exponentially with the number of degrees of freedom.\cite{stillexpgrows}
To correct for the uncompleteness of the available set of isomers, a
reweighting technique can be used.\cite{wales93,jcpquantum} The idea is to
perform a simulation at a reference temperature $T_0$, and to compute the
probability $p_\alpha$ of observing isomer $\alpha$.
The partition function $Z$ is approximated as a {\em partial} sum over
{\em weighted} isomers:
\begin{equation}
Z(\beta) \propto {\sum_\alpha}' g_\alpha n_\alpha Z_\alpha(\beta),
\label{eq:zreweight}
\end{equation}
with the partial character of the sum indicated by the prime symbol.
If we assume that the weights $\{ g_\alpha\}$ are independent of temperature,
the probability $p_\alpha(T_0)$ of finding isomer $\alpha$ at $T_0$ is
proportional to $g_\alpha n_\alpha Z_\alpha(\beta_0)$. Hence, $Z$ can be
expressed as
\begin{equation}
Z(\beta) \propto {\sum_\alpha}' p_\alpha(\beta_0) \frac{Z_\alpha(\beta)}
{Z_\alpha(\beta_0)}.
\label{eq:zreweight2}
\end{equation}
A classical simulation can be used to sample the minima and their respective
weights.\cite{jcpquantum} In the above equation, the denominators $Z_\alpha
(\beta_0)$ should then be replaced by their classical analogue. Even though
the resulting partition function (\ref{eq:zreweight2}) should not depend on the
classical or quantum nature of the simulation, some differences can be seen
when using semiclassical simulations including quantum corrections.\cite{prop}
This reweighting procedure has been shown to improve significantly the
predictions of the superposition method. In particular, Wales was able to
obtain a 'S-bend' in the kinetic temperature of
Ar$_{55}$,\cite{wales93} as seen previously in simulations.\cite{labastie}

The other important limitation of the method is the harmonic approximation used
to calculate $Z_\alpha(\beta)$. Several phenomenological anharmonic corrections
have been proposed in the litterature for classical systems.%
\cite{anhardoye,ball} In the case of quantum systems, the optimal
normal modes method of Cao and Voth\cite{voth} can be used to include
anharmonicity to some extent, but the self-consistency required makes it
computationally heavy for large sets of isomers. Building upon a recent effort
to include systematic corrections from the energy surface topography via
classical perturbation theory,\cite{jcpanhar} we can attempt to include
anharmonicities in the quantum partition functions $Z_\alpha$ in a more
rigorous fashion. Since the
classical corrections are also numerically demanding, we restrict ourselves
to low-order corrective terms. In order to make the reading easier, the details
of the perturbation expansion are given in the Appendix. The
superposition method not only provides the thermodynamical equilibrium
properties, and other quantities can be computed provided they can be
calculated for each minimum $\alpha$.\cite{prop} At thermal equilibrium, the
average value $\langle A\rangle$ of the global observable $A$ is the weighted
sum over minima:
\begin{equation}
\langle A\rangle (T) = \frac{\sum'_\alpha p_\alpha(T_0)Z_\alpha(T) A_\alpha(T)
/ Z_\alpha(T_0)}{\sum'_\alpha p_\alpha(T_0) Z_\alpha(T)/Z_\alpha(T_0)}.
\label{eq:aav}
\end{equation}

\section{Gaussian theory of absorption}
\label{sec:gaussian}

We assume that the cluster is in canonical equilibrium at temperature $T=1/k_B
\beta$, and that it undergoes an excitation by an infinitely narrow pulse at
frequency $\omega$ from its ground state potential surface $V_g({\bf R})$. The
normalized excitation spectrum ${\cal I}(\omega,\beta)$ is\cite{yan}
\begin{eqnarray}
&&{\cal I}(\omega,\beta) = \nonumber \\
&&\frac{\sum_i \sum_j \delta( \Delta E + E_{e_j} -
E_{g_i} - \hbar \omega) e^{-\beta E_{g_i}} | \langle e_j | \mu | g_i \rangle
|^2}{\sum_i \sum_j e^{-\beta E_{g_i}}| \langle e_j | \mu | g_i \rangle |^2},
\label{eq:intens0}
\end{eqnarray}
with $|g_i\rangle$ and $|e_j\rangle$ the eigenstates with eigenenergies
$E_{g_i}$, $E_{e_j}$ of the ground- and excited-state Hamiltonians $H_g$ and
$H_e$, respectively. The transition dipole moment $\mu_{ij}=|\langle e_j
| \mu | g_i \rangle|^2$ will be taken as constant in
the usual Condon approximation of vertical excitations. $\Delta E$ is the
energy difference between the bottom of the ground- and excited-state surfaces.
The absorption intensity can be expressed as the Fourier transform of the
imaginary-time correlation function:\cite{wp}
\begin{equation}
{\cal I}(\omega,\beta) = \frac{\hbar}{2\pi Z(\beta)}\int_{-\infty}^\infty
\chi(\tau,\beta)e^{-i\tau (\Delta E - \hbar \omega)} d\tau,
\label{eq:intens1}
\end{equation}
with
\begin{equation}
\chi(\tau,\beta) = {\rm Tr}\, \left[ e^{-i\tau H_e} e^{-(\beta - i\tau)H_g}
\right],
\label{eq:chi}
\end{equation}
and where $Z(\beta)={\rm Tr}\, e^{-\beta H_g}$ is the partition function.
Following Wadi and Pollak,\cite{wp} we now perform a short-time expansion
in the correlation function. This expansion should be valid at moderate
temperatures and for a large number of degrees of freedom. Indeed, in such
cases the combination of the numerous vibrational modes will significantly
dephase rapidly, and the correlation function will be nonzero only for short
times. Using this approximation, the absorption intensity can be calculated
analytically\cite{wp} as a Gaussian function:
\begin{equation}
{\cal I}(\omega,\beta) = \frac{\hbar}{\bar \omega \sqrt{2\pi}}\exp\left[
-\frac{1}{2}\left(\frac{\omega - \omega_0 - \Delta \omega}{\bar \omega}
\right)\right],
\label{eq:intensg}
\end{equation}
with $\omega_0 = E_0/\hbar$, $E_0$ being the vertical transition energy.
The line shift
$\Delta \omega$ and the width $\bar\omega$ are given as a function of the two
first moments of the shifted Hamiltonians difference $\Delta H = H_e - H_g -E_0
$ by\cite{wp}
\begin{equation}
\hbar \Delta \omega = \langle \Delta H \rangle,
\label{eq:domega0}
\end{equation}
and
\begin{equation}
\hbar \bar \omega = \langle \Delta H^2 \rangle - \langle \Delta H\rangle^2.
\label{eq:omegab0}
\end{equation}
We have noted here $\langle A\rangle$ the thermal average
of observable $A$ over the ground state potential surface $H_g$. We now write
the Hamiltonians $H_g$ and $H_e$ in harmonic (normal modes) coordinates ${\bf
Q}$:
\begin{eqnarray}
H_g &=& \frac{1}{2}\sum_i p_i^2 + \frac{1}{2} {\bf Q}^\dagger {\bf W}_g {\bf Q}
\label{eq:hg} \\
H_e &=& \frac{1}{2}\sum_i p_i^2 + {\bf G}^\dagger {\bf Q} + \frac{1}{2} {\bf
Q}^\dagger {\bf W}_e {\bf Q} + E_0.
\label{eq:he}
\end{eqnarray}
In these equations, ${\bf W}_g$ is the mass-weighted Hessian matrix of the
ground state at the lowest energy minimum ${\bf Q}_0$, which is assumed to
be located at the origin ${\bf 0}$ of the reference frame.
${\bf W}_g$ can be written as ${\bf W}_g = {\rm diag}\, (\omega_{g_i}^2)$, as
it is diagonal by construction.
${\bf W}_e = (w_e^{ij})$ is the second derivatives matrix of the excited
state surface at ${\bf Q}_0$. Lastly, ${\bf G}=\partial H_e({\bf Q}_0)/
\partial {\bf Q}$ is the excited state gradient at ${\bf Q}_0$. The matrix
${\bf W}_e$ is not assumed to be diagonal, and may include the so-called
Duschinskii rotations.\cite{duschinskii} It is then a simple matter of
algebra to obtain the
expressions for $\Delta \omega$ and $\bar\omega$. Using the notation
$\rho_i = \omega_{g_i}{\rm tanh}\, \beta \hbar \omega_{g_i}/2$, we find
\begin{eqnarray}
\Delta \omega &=& \frac{1}{4} \sum_i \frac{\omega_{g_i}^2 - w_e^{ii}}
{\rho_i} \label{eq:domega} \\
\bar\omega^2 &=& \frac{1}{8}\sum_i \left(\frac{\omega_{g_i}^2 - w_e^{ii}}
{\rho_i}\right)^2 + \frac{1}{8} \sum_{i\neq j} \frac{(w_e^{ij})^2}{
\rho_i\rho_j} + \frac{1}{2} \sum_i \frac{G_i^2}{\hbar \rho_i}.
\label{eq:omegab}
\end{eqnarray}
\begin{figure*}[htb]
\vbox to 7cm{
\includegraphics{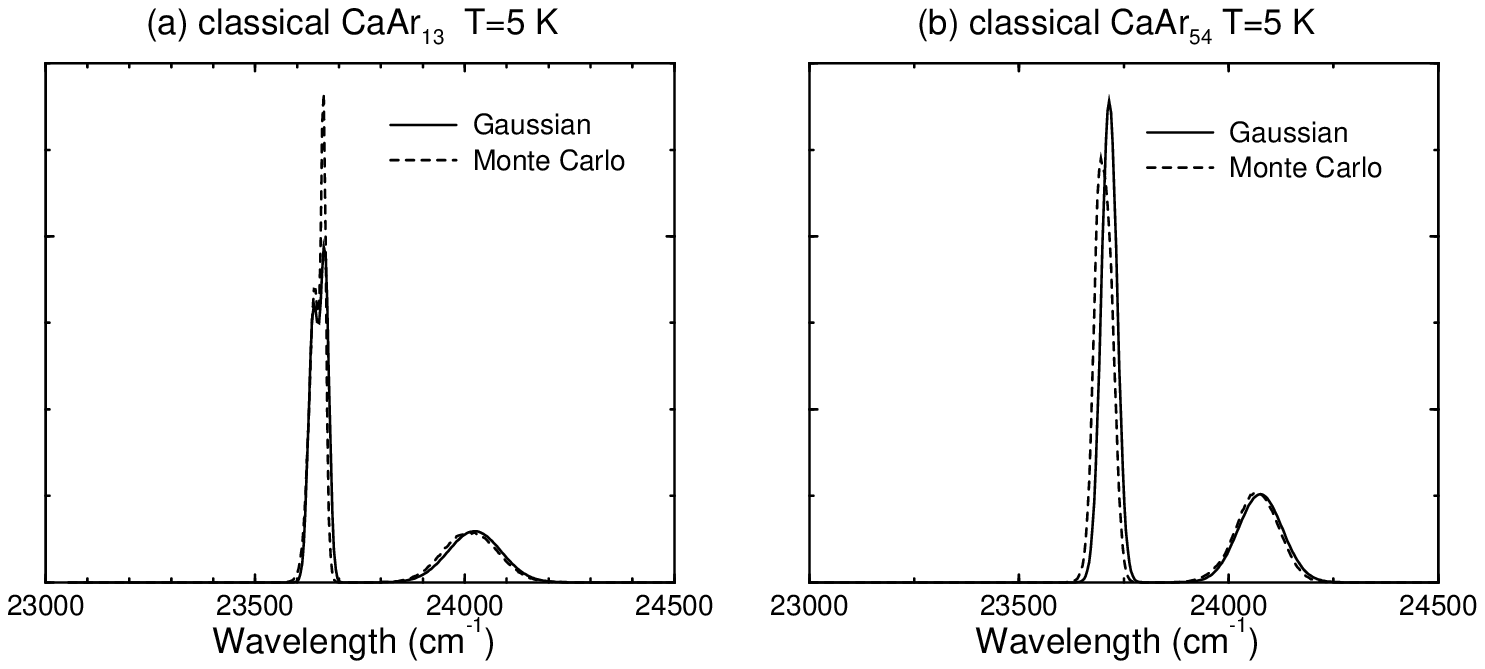}
\vfill}
\caption{Classical absorption spectra of the $4s^2\to 4s4p$ transition in
(a) CaAr$_{13}$ and (b) CaAr$_{54}$ from Monte Carlo simulations (dashed lines)
and the Gaussian theory (solid lines) at $T=5$~K.}
\label{fig:classpectra}
\end{figure*}
Apart from the different notations, the above expressions differ from the
results by Wadi and Pollak by the second term in the right-hand side of
Eq.~(\ref{eq:omegab}), which was
neglected by these authors.\cite{wp} Although Eqn.~(\ref{eq:domega}) and
(\ref{eq:omegab}) are only valid for polyatomic, harmonic systems, anharmonic
corrections can be carried out in the very same way as the perturbative
expansion mentioned in the previous section and detailed in the Appendix.
However, the present expressions are already the outcome of an approximation
in the correlation function, hence we did not attempt to include such
corrections here. In addition, their estimation would require the computation
of higher order derivatives of the excited states energy surfaces, which
will be a heavy numerical task, even for for DIM potentials.

We have tested the Gaussian theory on small CaAr$_n$ clusters. The ground-
and excited-state Hamiltonians have been fully described previously.%
\cite{epjd} They are based on the Diatomic-In-Molecules (DIM) formalism,%
\cite{dim} and were parameterized on {\em ab initio} calculations.%
\cite{abinit} To give a better account of experimental properties of the
CaAr diatomics, we have slightly modified the parameters with respect to our
previous works.\cite{epjd,prop} The Ca--Ar ground state pairwise potential has
been scaled by 0.85, and the excited state surfaces have been shifted by
+305~cm$^{-1}$ to fit the $4s4p$ experimental transition in calcium.
As a consequence, some global minima may have changed, as in
CaAr$_{37}$.

In the present work, we focus on the $4s^2\to 4s4p$
transitions, which have received some experimental attention.\cite{epjd}
The DIM Hamiltonian provides the energy and numerical derivatives, as well as
the electronic transition dipole moment, for any atomic configuration
${\bf R}$. It can thus be used to
simulate absolute absorption spectra. The ground state vibrational eigenmodes
and eigenfrequencies $\{\omega_{g_i}\}$ were obtained from the diagonalization
of the Hessian matrix whose elements were computed from their analytical
expressions. In both the classical and quantum cases, the absolute spectrum
is calculated as the sum over the three possible excited states.
For the Gaussian theory, the three gaussian curves are weighted by the
square of the transition dipole moment at the equilibrium ground state
geometry. 

First, we have calculated the classical spectrum from the Gaussian
theory in the classical limit $\hbar\to 0$, and from classical Metropolis
Monte Carlo simulations at low temperature, using $10^6$ cycles
following $2\times 10^5$ equilibration cycles. The spectra of CaAr$_{13}$ and
CaAr$_{54}$ at $T=5$~K are displayed in Fig.~\ref{fig:classpectra}. For
comparison, we have also plotted the predictions of the Gaussian theory in
the classical limit $\hbar\to 0$. In this limit, the line shift and line width
tend to their classical values $(\hbar \Delta \omega)_c$ and $(\hbar \bar
\omega)^2_c$ given respectively by
\begin{eqnarray}
(\hbar \Delta \omega)_c &=&\frac{k_BT}{2}\sum_i \frac{\omega_{g_i}^2 -
w_e^{ii}}{\omega_{g_i}^2}, \label{eq:cdomega} \\
(\hbar \bar\omega)_c^2 &=& \frac{(k_BT)^2}{2}\sum_i \left(\frac{\omega_{g_i}^2
- w_e^{ii}}{\omega_{g_i}^2}\right)^2 \nonumber \\
&&+ \frac{(k_BT)^2}
{2}\sum_{i\neq j} \frac{(w_e^{ij})^2}{\omega_{g_i}\omega_{g_j}}
+ k_BT \sum_i \frac{G_i^2}{\omega_{g_i}}.
\label{eq:comegab}
\end{eqnarray}
The reliability of our comparison was checked by ensuring
that only a single isomer was visited during the Metropolis MC simulation.
As can be seen in Fig.~\ref{fig:classpectra}, the accuracy of the Gaussian
theory is very good at such low temperatures. The global minimum geometry of
CaAr$_{13}$ has two spectral lines located at 23658 (2 degenerate lines) and
24034~cm$^{-1}$, respectively. Due to its finite temperature, the cluster
loses its instantaneous symmetry and the degeneracy is lifted. This explains
why three distinct lines are seen in Fig.~\ref{fig:classpectra}(a). In
CaAr$_{54}$ the Gaussian theory of absorption remains in good agreement with
Monte Carlo data. This cluster has again 2 degenerate sharp lines at
23715~cm$^{-1}$, and one broad line at 24080~cm$^{-1}$. This time, the
degeneracy is no longer lifted by temperature, and the two lines still merge
into a single peak. In the classical theory
the absorption intensity can also be obtained analytically by integrating Eq.~%
(\ref{eq:intens1}) under the Condon assumption without performing the
short-time expansion. Therefore, in the classical case, the differences between
the calculated and simulated spectra in Fig.~\ref{fig:classpectra} can be
attributed to anharmonicities only.

\begin{figure}[htb]
\vbox to 6.5cm{
\includegraphics{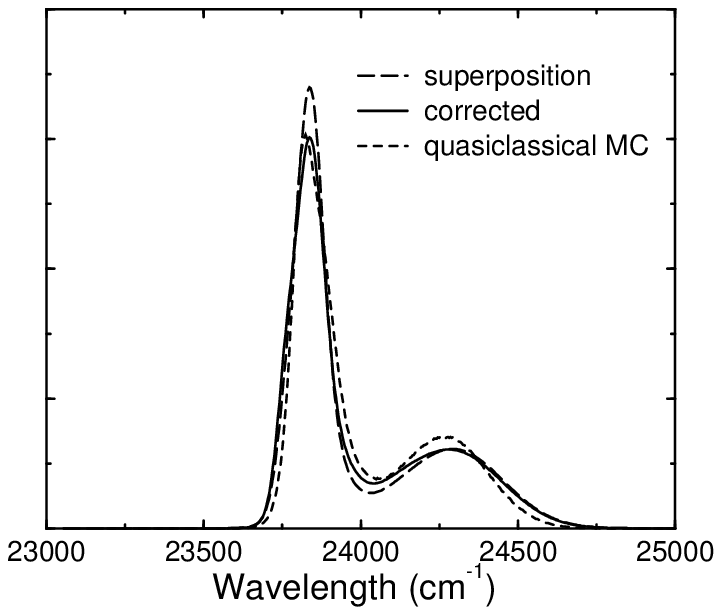}
\vfill}
\caption{Absorption spectra of CaAr$_{37}$ at $T=25$~K from quasiclassical
Monte Carlo simulations (dashed line) and the quantum superposition method
in the harmonic approximation (long dashed line) or with anharmonic
perturbative corrections (solid line).}
\label{fig:caar37}
\end{figure}
We now turn to a more complicated situation, namely CaAr$_{37}$ in the quantum
regime, and at moderate temperature $T=25$~K. We will show below that this
temperature is located just above the melting point of this cluster. Many
isomers are present, and this will provide a test of the superposition
approximation. In the simulation, quantum effects have
been included in a quasiclassical, corrective way by adding
the corrective Feynman-Hibbs effective potential\cite{fh} to the ground state
energy. This potential was shown to give good results on argon clusters with
respect to the more accurate Fourier-path-integral Monte Carlo simulation of
Neirotti and coworkers,\cite{nfd} especially when compared with the
alternative Wigner-Kirkwood corrections.\cite{wk} Since it is quasiclassical
in nature, it cannot be used in the low temperature regime $T\lesssim 5$~K
where only few isomers are present, and path-integral simulations would have
to be carried out instead.
Because the superposition method does not assume any barrier nor pathway
between the isomers, it yields supposedly ergodic data within the set of
available minima. Therefore a reliable comparison
with simulation results must also correct for possible broken ergodicity, and
for this we have used parallel tempering Monte Carlo.\cite{ptmc} Also, since
the effective potential explicitely depends on temperature, the probability of
accepting an exchange between adjacent trajectories should be modified
accordingly.\cite{telec}

Concerning the superposition approximation, and to quantify the importance
of perturbative corrections in the partition function, we have
separately considered zero-th order harmonic as well as second-order,
anharmonic-corrected partition functions with the same database of isomers.
The absorption spectra obtained with the Gaussian theory in the superposition
approach and with quasiclassical
MC simulations are represented in Fig.~\ref{fig:caar37}.
The lowest energy isomer of this cluster, including the zero-point energy
contribution, has three distinct spectral lines for the $4s^2\to4s4p$
transition, namely at 23840, 23970, and 24260~cm$^{-1}$. As can be seen from
Fig.~\ref{fig:caar37}, none of these lines remain visible as peaks.
We observe a very
good agreement between the present theory and on-the-fly simulation,
indicating that the relative weights of the isomers are correct.
However,
at this point, we cannot tell whether the visible shifts and broadenings
are an effect of temperature only, or of the different spectroscopic
signatures of the other isomers present. 
Some shifts in the peaks remain slightly
approximate in the superposition approach, and the second-order corrected
weights only marginally improve the results. At 25~K, quantum effects turn out
to be rather small on the line width and shape, but they may be more important
on the weights of the isomers, as will be seen below. Therefore, from
Fig~\ref{fig:caar37} we see that anharmonic corrections to the isomers weights
have a relatively weak influence for quasiclassical clusters. More
information on the importance of these corrections will be gained by looking at
the thermodynamical curves in the next section.

\begin{figure*}[htb]
\vbox to 7cm{
\includegraphics{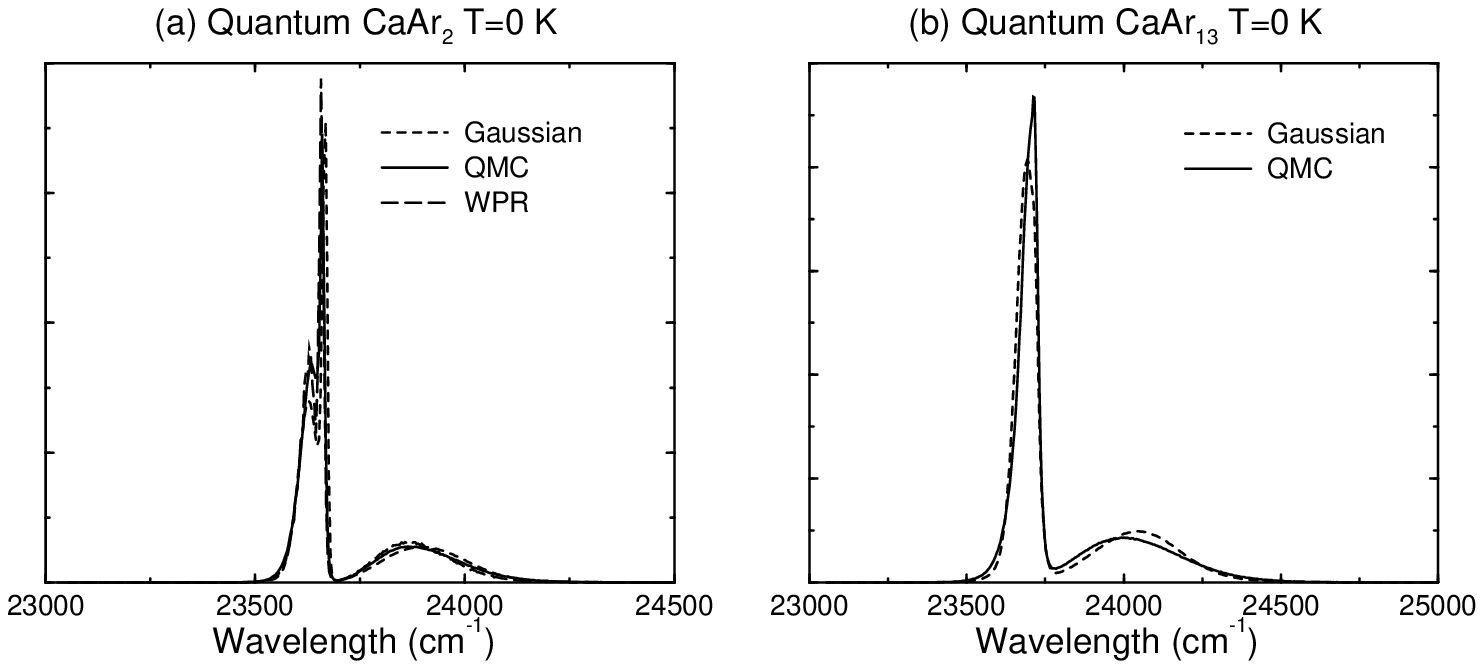}
\vfill}
\caption{Absorption spectra of (a) CaAr$_2$ and (b) CaAr$_{13}$ at $T=0$~K from
diffusion Monte Carlo simulations (solid lines) and the Gaussian theory
(dashed lines). For CaAr$_2$, we have also represented the results of
wavepacket relaxation calculations (long dashed line).}
\label{fig:qcaar}
\end{figure*}
The Gaussian theory presented in this section can be extended to purely
quantum systems at $T=0$, for which delocalization is only due to the quantum
nature of the vibrational modes, but not to an actual atomic displacement
induced by a finite energy deposit. In this purpose, the Boltzmann weights in
Eqn.~(\ref{eq:intens0}) and (\ref{eq:intens1}) are replaced by the probability
density of finding the system in its ground state at configuration ${\bf R}$
as the square of the wavefunction. In the harmonic approximation, this
probability density is a multidimensional Gaussian function, and the integrated
absorption spectrum is found to be Gaussian with line shape and width given by
the $\beta\to\infty$ limits of Eqn.~(\ref{eq:domega}) and (\ref{eq:omegab}):
\begin{eqnarray}
(\Delta\omega)_q &=&\frac{1}{4}\sum_i \frac{\omega_{g_i}^2 - \omega_e^{ii}}
{\omega_{g_i}} \label{eq:qdomega} \\
(\bar\omega^2)_q &=& \frac{1}{8}\sum_i \left( \frac{\omega_{g_i}^2 -
\omega_e^{ii}}{\omega_{g_i}}\right)^2 \nonumber \\
&& + \frac{1}{8}\sum_{i\neq j}\frac{
\omega_e^{ij}}{\omega_{g_i}\omega_{g_j}} + \frac{1}{2}\sum_i \frac{G_i^2}
{\hbar \omega_{g_i}}. \label{eq:qomegab}
\end{eqnarray}
The behavior of the Gaussian theory in the $T=0$ quantum case has been tested
on CaAr$_2$ and CaAr$_{13}$ using simple diffusion Monte Carlo (DMC) in
internal coordinates (without global
rotation and translation), and without importance sampling.
The DMC calculations were carried out with 1000 replicas, and the
wave functions were averaged over 10$^5$ steps after 10$^5$ initial
equilibration steps.
In addition, for CaAr$_2$, the vibrational ground state wavefunction was also 
computed numerically on grids, using quantum wavepacket relaxation (WPR)
method (imaginary time propagation).\cite{kos86:223}
The propagation was performed for zero total angular momentum,
representing the wavepacket by an harmonic oscillator 
Discrete Variable Representation (DVR)
for each of the two Ca--Ar bond coordinates, and by a Legendre DVR 
for the bending angle $\widehat{\rm Ar Ca Ar}$.
The primitive basis for representing one bond coordinate
includes 10 harmonic oscillator eigenfunctions.
Associated Legendre functions with $j<60$ and $m=0$ were used
for representing the bending mode. In both the DMC and WPR methods, which
are both supposed to give essentially exact results, the
absorption spectrum was calculated assuming vertical transitions from the
quantum sample of geometries.

In Fig.~\ref{fig:qcaar} the absorption spectra calculated with these various
methods are compared. For the small, triatomic cluster, the three methods
give very similar peaks. In its equilibrium geometry, CaAr$_2$ has three
nondegenerate spectroscopic lines at 23621, 23670, and 23918~cm$^{-1}$,
respectively. Quantum delocalization of the vibrational modes leads to a rather
small shift of these lines, and to a significant broadening.
For CaAr$_{13}$, the simulated peaks are also in
good agreement with the Gaussian prediction, but we notice that some asymmetry
is visible on the ``exact'' spectrum from diffusion Monte Carlo. These
effects are clearly beyond a Gaussian description, and are due to anharmonic
effects. These effects are relatively weak for the rather heavy calcium-doped
argon clusters but may become more important for neon or helium clusters.

The above results suggest that the Gaussian theory of photoabsorption
developed by Wadi and Pollak\cite{wp} upon the formal results of Yan and
Mukamel\cite{yan} can give quite accurate results in both the classical
and quantum regimes at low and moderate temperatures.

\section{Spectroscopic signature of phase changes}
\label{sec:appl}

\begin{figure*}[htb]
\vbox to 12cm{
\includegraphics{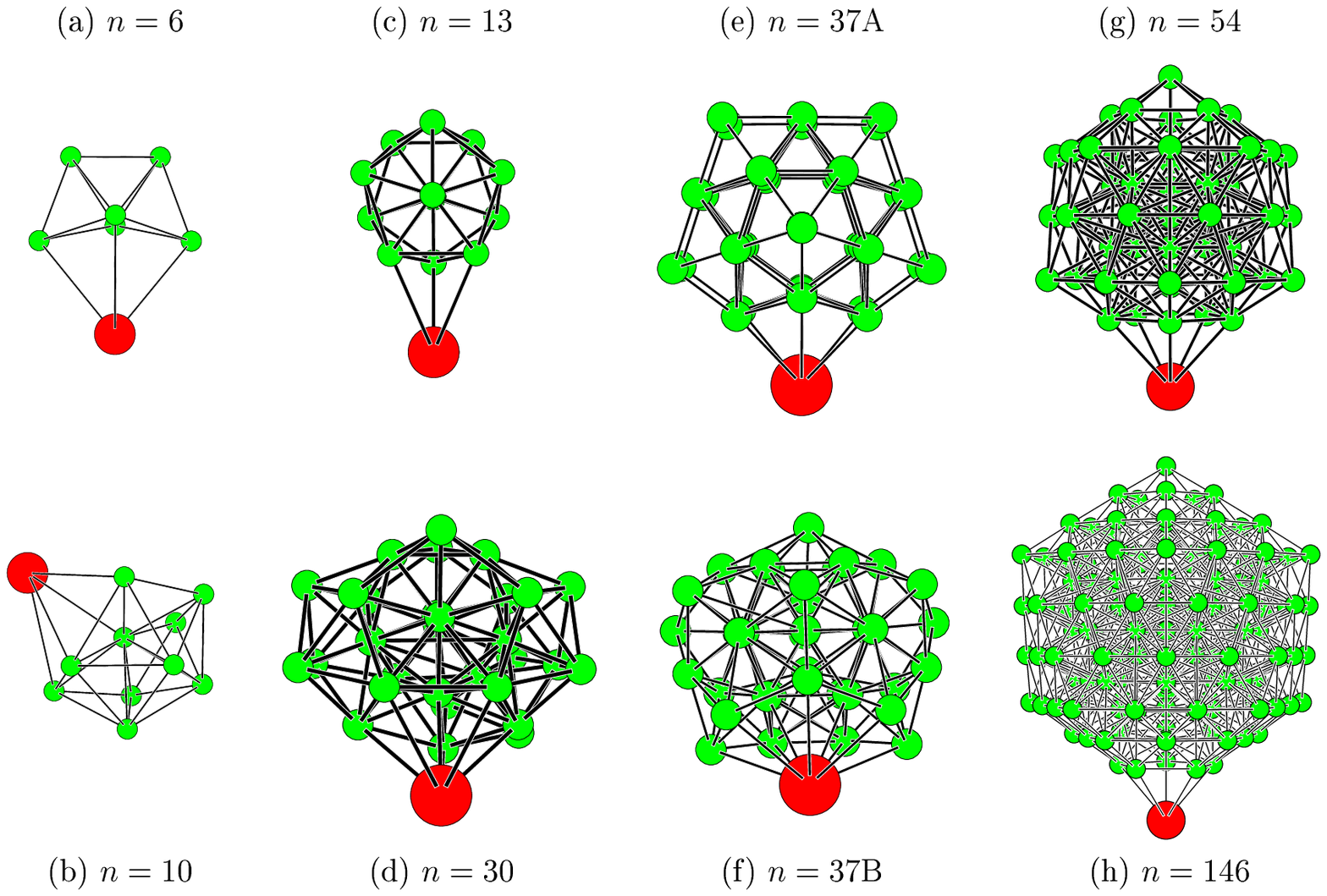}
\vfill}
\caption{Global minima of CaAr$_n$ found by the basin-hopping algorithm after
including the harmonic zero-point energy. (a) $n=6$; (b) $n=10$; (c) $n=13$;
(d) $n=30$; (e) $n=37$; (f) second lowest energy minimum of CaAr$_{37}$; (g)
$n=54$; and (h) $n=146$.}
\label{fig:struct}
\end{figure*}

The Gaussian theory described in the previous Section will now be used with
the superposition approach to investigate the influence of temperature on the
absorption spectra in context with isomerization phenomena.

For the two smallest clusters chosen, CaAr$_6$ and CaAr$_{10}$,
we performed extensive sampling of the ground state PES, and we
believe that nearly all stable isomers have been identified. 
To apply the superposition method to larger CaAr$_n$ clusters, the
reweighting procedure from simulations must be used. 
Quasiclassical parallel tempering Monte Carlo simulations of $10^7+2\times
10^6$ equilibration cycles were used, for a set of temperatures equally
spaced by 2.5~K in the range 7.5~K$\leq T\leq 60$~K for $n\leq 54$, and
15~K$\leq T\leq 75$~K for CaAr$_{146}$. The isomers were found by quenching
the 50~K trajectory for all sizes except CaAr$_{37}$, for which the 25~K
trajectory was used, and CaAr$_{146}$, quenched at 60~K. Local minimizations
were performed every 500 cycles. To prevent
evaporation, a rigid spherical container was used in the MC simulation.%
\cite{lba,labastie} The radii of the containers and the number of different
isomers obtained from the quenches are given in Table~\ref{table} for the set
of sizes chosen. We have also indicated the point group and energy of the
global minima after including the zero-point
harmonic contribution. These lowest-energy structures were generated from the
basin-hopping or Monte Carlo + minimization algorithm.\cite{bh} They are shown
in Fig.~\ref{fig:struct}. In most cases, the global
minimum of CaAr$_n$ is very similar to the one of Ar$_{n+1}$, with one argon
atom substituted by calcium. Due to the lower binding energy and larger
equilibrium distance of CaAr with respect to Ar$_2$,\cite{abinit} calcium
is always found on the surface of
\begin{table*}[htb]
\caption{Properties of the CaAr$_n$ clusters simulated in this work. The energy
and point group refer to the lowest energy minima, including
the harmonic zero-point contribution.}
\label{table}
\begin{tabular}{l|c|c|c|c|c}
\colrule
Cluster & Energy & Point group 
& Container radius & Quenching temperature & Number of isomers \\
size & (eV/atom) & & (\AA) & (K) & \\
\colrule
CaAr$_6$     & 0.02388 & C$_{2v}$ &  6.348 & -- &   12 \\
CaAr$_{10}$  & 0.03192 &   C$_s$  &  7.935 & -- &  391 \\
CaAr$_{13}$  & 0.03721 & C$_{3v}$ &  7.935 & 50 &  717 \\
CaAr$_{30}$  & 0.04808 &   C$_1$  & 11.638 & 50 & 2347 \\
CaAr$_{37}$  & 0.05044 & C$_{4v}$ & 11.638 & 25 & 1444 \\
CaAr$_{54}$  & 0.05690 & C$_{5v}$ & 13.225 & 50 & 2563 \\
CaAr$_{146}$ & 0.06695 & C$_{5v}$ & 15.870 & 60 & 1690 \\
\colrule
\end{tabular}
\end{table*}
the remaining argon cluster. The nearly identical masses of calcium and argon
atoms also induces relatively small changes in the vibrational properties, and
the similarities between CaAr$_n$ and Ar$_{n+1}$ remain when including the
zero-point corrections. However, in the case of CaAr$_{37}$, we find the
somewhat surprising result that the global minimum geometry is decahedral, and
remains so after including the harmonic zero point energy contribution.
Slightly higher in energy, several icosahedral structures are found, either in
the form of uncomplete multilayer icosahedra (Mackay type) or as polyicosahedra
(anti-Mackay type). As in Ar$_{38}$, quantum delocalization favors anti-Mackay
structures over Mackay shapes,\cite{jcpquantum} yet a Mackay structure is found
as the second isomer in both the classical and quantum regimes. This isomer is
represented in Fig.~\ref{fig:struct}(f). Additionally, the distorted truncated
octahedron geometry lies above, but very close, to these isomers. As a
consequence, the energy landscape of this particular cluster reveals interesting
features, which will be investigated more deeply in the following article of
this series.\cite{papdjw} This result is also consistent with the richer
thermodynamical behavior generally observed in mixed van der Waals clusters 
with respect to homogeneous clusters.\cite{mixedvdw1,mixedvdw2}

Therefore, even though CaAr$_n$ and Ar$_{n+1}$ clusters share many structural
features, they also exhibit significant and unexpected differences, which could
affect the thermodynamics and dynamics in large extents.

\subsection{\boldmath Isomerization in CaAr$_6$ and CaAr$_{10}$\unboldmath}

\begin{figure*}[htb]
\vbox to 13.8cm{
\includegraphics{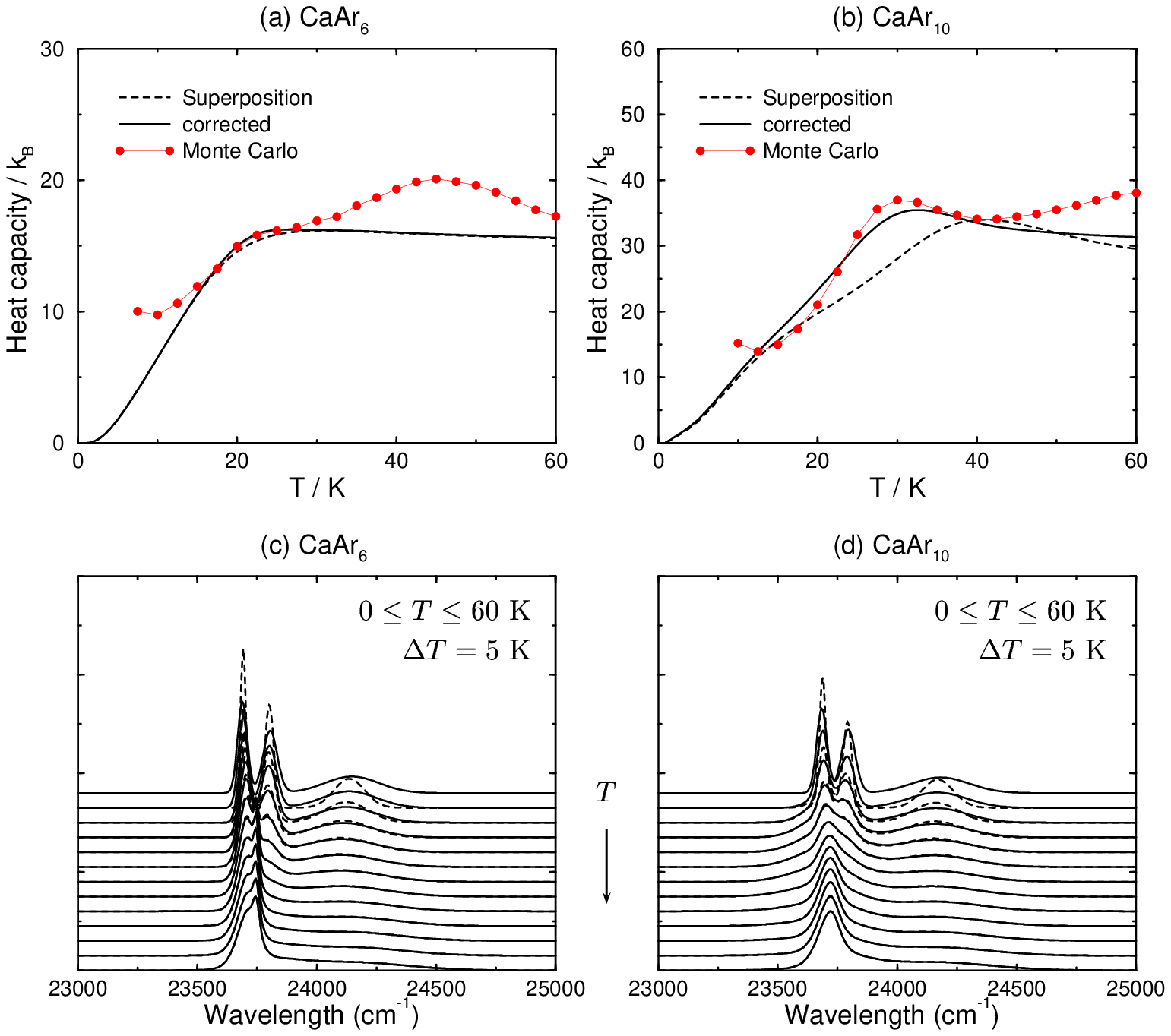}
\vfill}
\caption{Heat capacities computed from quasiclassical Monte Carlo simulations
(full circles) and from the superposition method with harmonic (dashed lines)
or anharmonic (solid lines) partition functions for (a) CaAr$_6$ and (b)
CaAr$_{10}$. Also represented are the finite-temperature absorption spectra
calculated from the superposition method and the Gaussian theory of absorption
in the quantum (solid lines) and classical (dashed lines) regimes for (c)
CaAr$_6$ and (d) CaAr$_{10}$.}
\label{fig:small}
\end{figure*}

The lowest energy minimum of CaAr$_6$ is a distorted pentagonal bipyramid.
The heat capacity of this cluster, computed from the superposition approach
with or without anharmonic perturbative corrections,
is represented in Fig.~\ref{fig:small}(a). For comparison,
we have also plotted on the same curves the direct results of parallel
tempering Monte Carlo with the Feynman-Hibbs effective potential.\cite{fh}
As is well known for this potential,\cite{jcpquantum} the canonical heat
capacity diverges upwards at low temperatures.
Due to its small size, this cluster shows a nearly featureless caloric curve,
and a flat heat capacity (except the low temperature increase) in
Fig.~\ref{fig:small}(a). The small hump at $T\sim 20$--25~K is indicative of
isomerizations, and is followed by another bump near 45~K in the MC simulation.
This latter broad peak is in fact spurious, as it comes from the dissociation
of the weakly bound Ca atom. Previous works\cite{nordiek,calvoc60} have
identified this peak as the signature of the liquid-vapor transition, from
its dependence on the container radius. Here it would have been difficult to
employ a radius smaller than 6\AA, because the motion of argon atoms would have
been hindered. The 25~K hump is barely visible, because it is hidden and
smoothed out by
quantum effects. The agreement between the superposition approximation and the
quasiclassical MC simulations is very good except for the extra liquid-vapor
bump. This feature is obviously missing from the assumptions made in the
superposition method. The quantitative effect of perturbative corrections is
small for this cluster.

CaAr$_{10}$ has the ground state of an uncomplete icosahedron. Some isomers
involving different locations of the calcium atom lie not far above the global
minimum, resulting in a shoulder in the heat capacity near 10--15~K in
Fig.~\ref{fig:small}(b). The main peak near 30~K is due to a more complete
rearrangement of argon atoms, and is reproduced quantitatively by the
superposition approach after including perturbative corrections in the
partition function.

\begin{figure*}[htb]
\vbox to 13.8cm{
\includegraphics{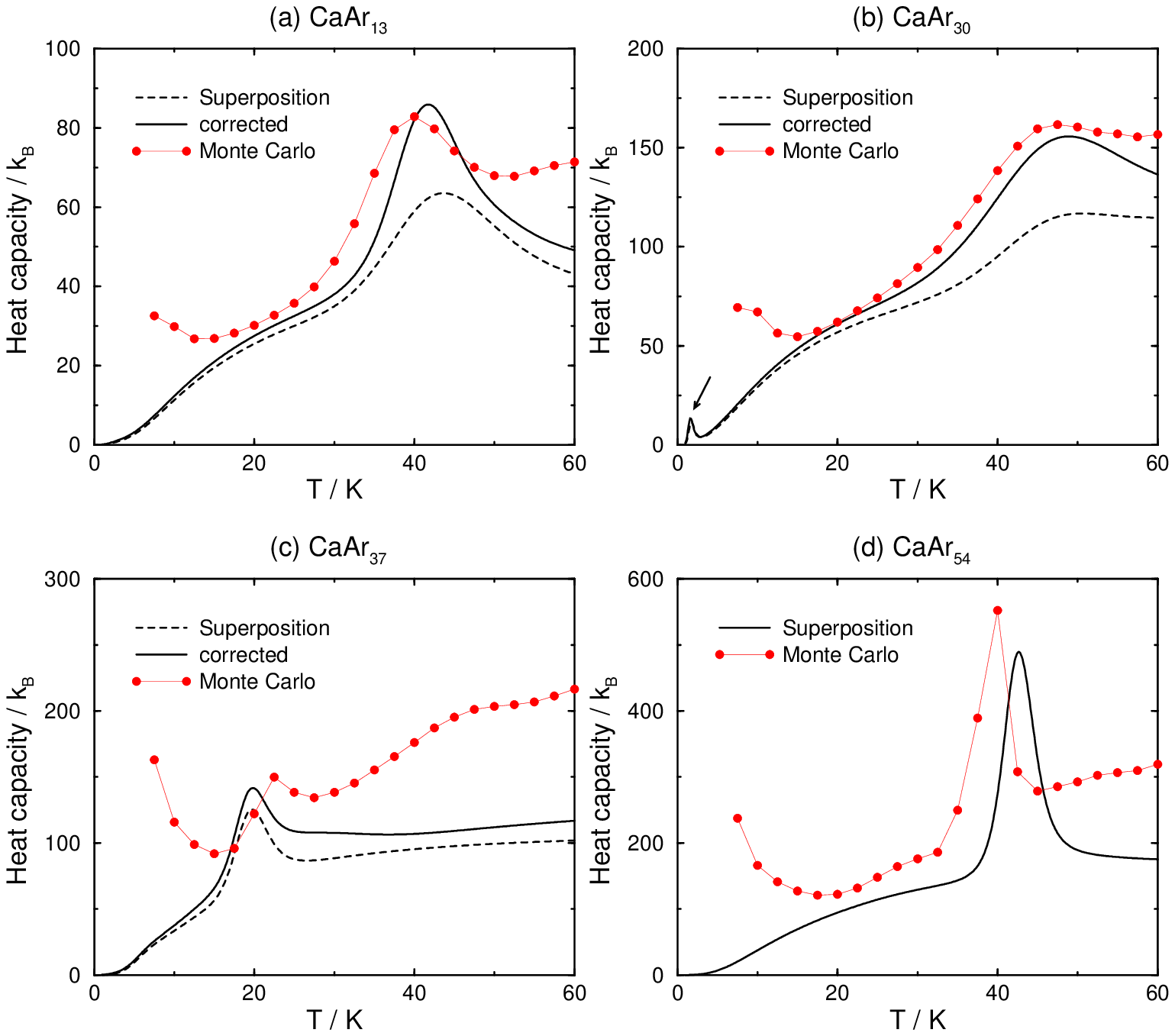}
\vfill}
\caption{Heat capacities of CaAr$_n$ clusters. (a) $n=13$; (b) $n=30$; (c)
$n=37$; and (d) $n=54$. The notations are the same as
in Fig.\protect~\ref{fig:small}.}
\label{fig:cv}
\end{figure*}

The absorption spectra of these two clusters are represented in
Fig.~\ref{fig:small}(c) and \ref{fig:small}(d) as a function of temperature.
In each of these figures, the classical spectra obtained from the limit
$\hbar\to 0$ has been superimposed to the quantum spectra.
For the two sizes, the patterns are rather
similar: the spectrum varies from a clear three-peak structure at low
temperature towards a mostly single peak shape at high temperatures. The
transition occurs roughly at 20~K in CaAr$_6$, and near 30~K in CaAr$_{10}$.
In the case of CaAr$_6$ the high temperature main peak appears made of two
smaller peaks, in agreement with simulations.\cite{isspic} From a general point
of view, quantum effects mainly consist in a broadening of the peaks. As
expected, they get smaller and smaller as temperature increases, and becomes
barely visible above 25~K.

Therefore, in these two small clusters, absorption spectroscopy seems to be
an effective mean of detecting isomerization events, as the range of variation
in the spectrum approximately matches the width of the heat capacity changes.
As cluster size grows, one can thus expect to see sharper transitions, and to
possibly use the calcium chromophore as a probe of the argon cluster melting
point.

\subsection{\boldmath Phase changes in CaAr$_n$, $13\leq n\leq 54$\unboldmath}

We have chosen a set of sizes in the range $13\leq n\leq 54$, spanning between
1 and 2 icosahedral layers, and which are likely to exhibit significant finite
size effects. We first discuss the thermodynamical behavior of these
clusters. The heat capacities obtained using the superposition method, with
harmonic or second-order corrected anharmonic partition functions, are
represented in Fig.~\ref{fig:cv} for the clusters corresponding to $n=13$,
30, 37, and 54. In the case of the larger cluster CaAr$_{54}$, and also for
CaAr$_{146}$, it was not possible to include all the required anharmonic
corrections due to the too large computational requirements. 
Since they are not the main focus of the
present paper, these thermodynamical data will be only briefly discussed in
order to guide our interpretation of the following absorption spectra.

\begin{figure*}[htb]
\vbox to 7cm{
\includegraphics{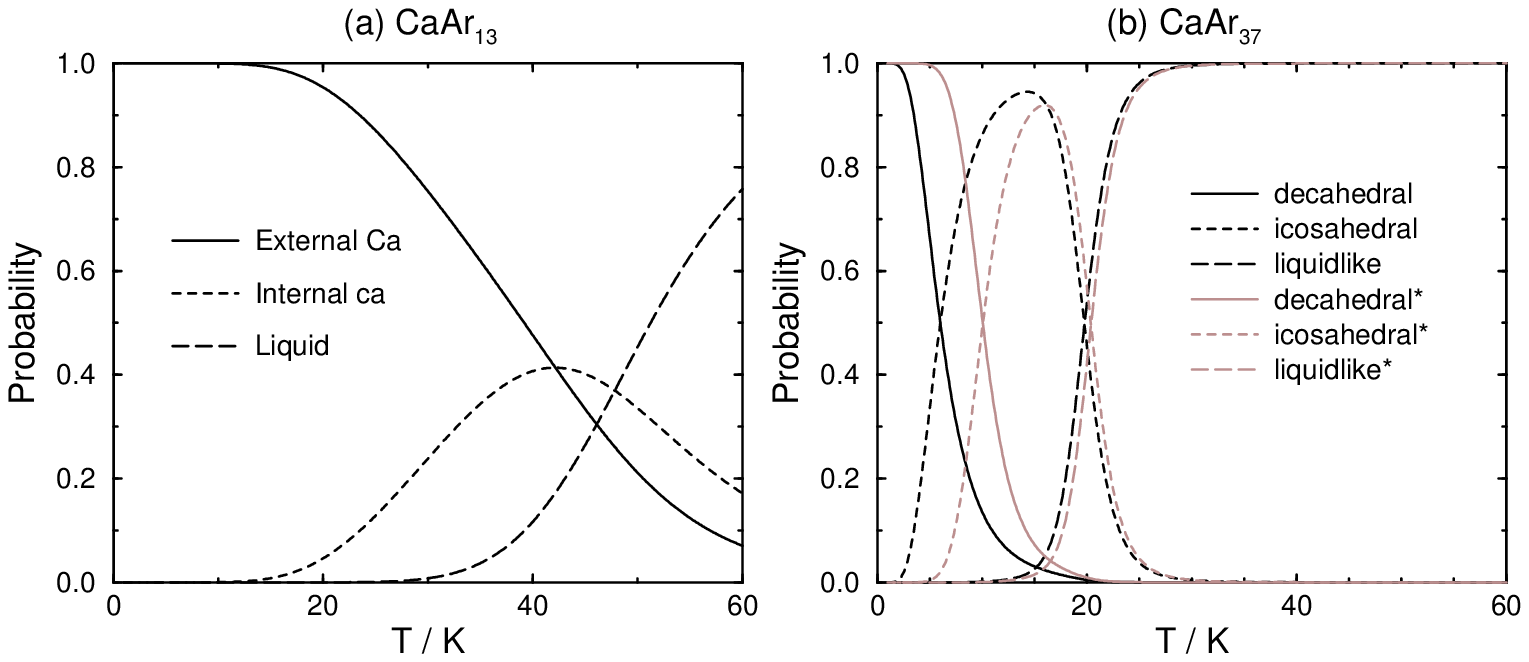}
\vfill}
\caption{(a) Probability to find the CaAr$_{13}$ cluster with the calcium atom
in a capping position (solid line), inside the icosahedral shell (dashed line),
or in the liquidlike state (long dashed line). (b) Probability to find the
CaAr$_{37}$ cluster in its decahedral ground state (black solid line), in a
Mackay-type icosahedral isomer (black dashed line), or in any other isomer
(black long dashed line). The corresponding classical curves are also plotted
as grey lines.}
\label{fig:isom}
\end{figure*}

In general, we find again
a rather good agreement between the quantum superposition
results and the Monte Carlo simulations for all four sizes. As expected,
anharmonic corrections do have a positive quantitative effect by bringing
the superposition data closer to the simulation results. However, no new
qualitative feature is visible when they are taken into account. All heat
capacities exhibit a main peak indicative of the solidlike-liquidlike phase
change.\cite{labastie} This peak is located near 40--45~K for CaAr$_{13}$,
CaAr$_{30}$ and CaAr$_{54}$, and near 20~K for CaAr$_{37}$. This lower melting
point is similar to what is observed in the 38-atom argon cluster described
by the Lennard-Jones potential.\cite{ar38doye,ar38calvo}
In addition to this main peak, extra humps or even peaks can be observed at
lower temperatures in CaAr$_{30}$ and CaAr$_{37}$. These features are
indicative of preliminary isomerization resulting from solid-solid transitions
in these clusters.\cite{doyecalvo} Again, they bear similarities with the
known finite temperature behavior of LJ clusters with the same size.%
\cite{ar38doye,ar38calvo,ar31,frantz} In CaAr$_{13}$, the quantum character of
the specific heat hides again a shoulder in the classical curve, as is the
case in CaAr$_6$. This cluster happens
to have a much larger number of stable isomers than Ar$_{14}$. In particular
the capped icosahedron can be found with the calcium atom either in adatom
location (lowest energy structure) or inside the icosahedral shell, resulting
in a set of 4 different isomers slightly higher in energy. Structures
not based on the capped icosahedron, or with the calcium atom in
the center of the icosahedron, are less stable than these five geometries.
Therefore the energy spectrum of local minima can be put into three sets, one
being the global minimum only, one being the other capped icosahedral isomers
with calcium in the icosahedral shell, and all other isomers not build on the
capped icosahedron motif. Because of these three sets, the melting process
involves essentially
two steps. In Fig.~\ref{fig:isom}(a) we plot the probabilities of finding the
CaAr$_{13}$ cluster in its ground state, or in one of the 4 subsequent isomers,
or in the remaining, liquid configurations. These probabilities are computed
using the anharmonic corrected quantum partition functions. They display a
clear two-step isomerization pattern. In the first step, at about 30~K, the
calcium atom is able to leave its capping site in favor of one of the twelve
icosahedral sites. Then, above 40~K, more disordered structures become
thermally available the melting peak takes place. 

In CaAr$_{30}$ the preliminary solid-solid transition near $2.5$~K involves
polyicosahedral and multilayer icosahedral isomers, as in
Ar$_{31}$.\cite{jcpquantum,ar31}
Even though CaAr$_{37}$ does not have the same global minimum as Ar$_{38}$, a
similar thermodynamic behavior occurs. We have
calculated from the harmonic superposition approximation the probabilities of
finding the cluster either in its decahedral ground state isomer, in
its (Mackay-type)
icosahedral isomers, or in any of the remaining isomers. The variation of these
probabilities with increasing temperature are displayed in
Fig.~\ref{fig:isom}(b). They show that the transition at about
5~K is due to a global decahedral-icosahedral rearrangement, and bears some
similarity with the (C$_{60}$)$_{14}$ cluster.\cite{prlcomment}
The anti-Mackay and truncated octahedral structures are only marginally
populated at thermal equilibrium, and the main heat capacity peak is associated
with the emergence of a large number of disordered configurations. Hence it
can be considered as the signature of melting. For comparison, we have also
plotted in Fig.~\ref{fig:isom}(b) the corresponding curves in the classical
limit. While the global behavior remains essentially unchanged, the
preliminary transition occurs at a higher temperature, near 10~K.

\begin{figure*}[htb]
\vbox to 13.8cm{
\includegraphics{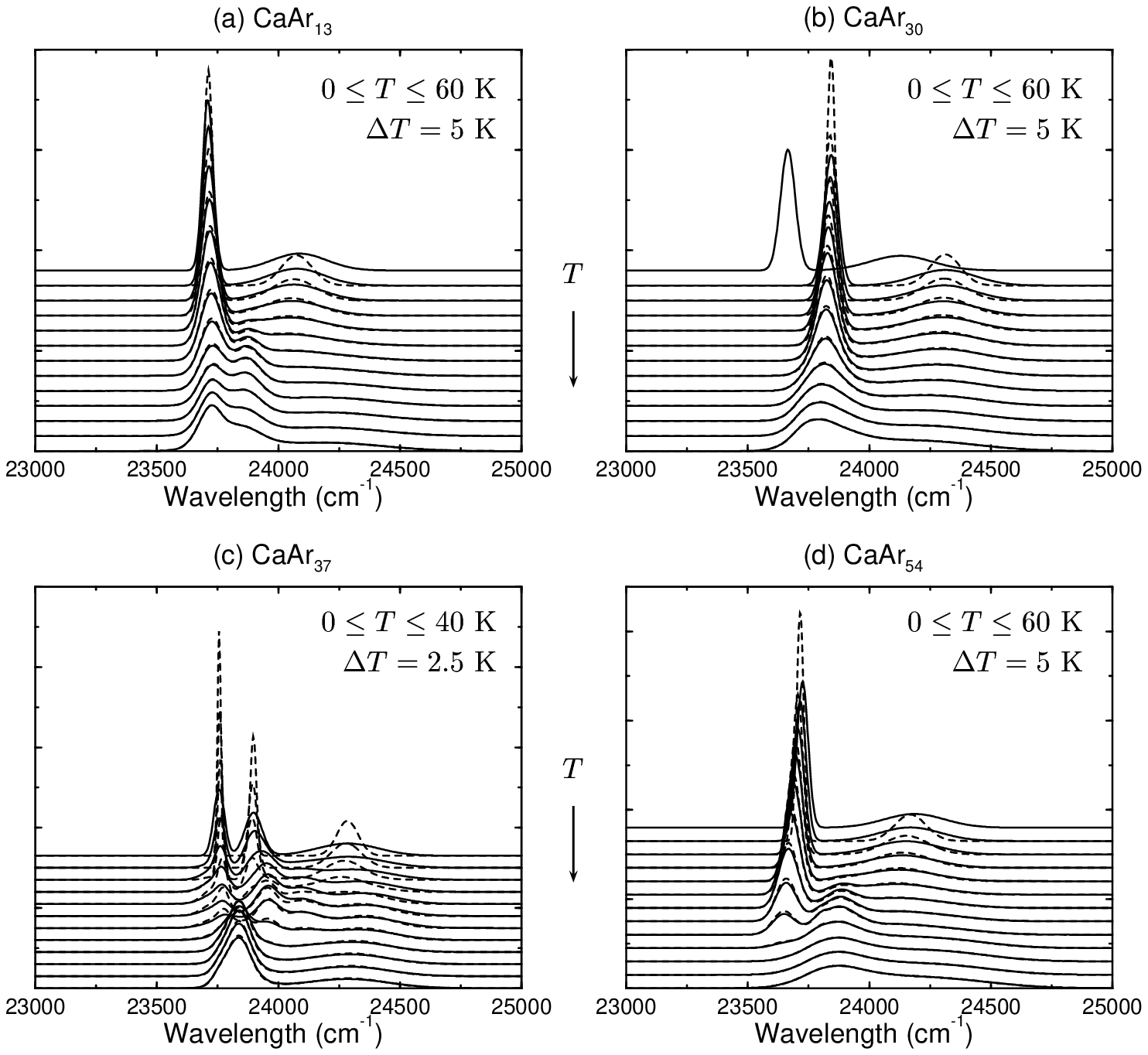}
\vfill}
\caption{Absorption spectra of the CaAr$_n$ clusters corresponding to the
heat capacities of Fig.~\protect\ref{fig:cv}. (a) $n=13$; (b) $n=30$; (c)
$n=37$; and (d) $n=54$. The notations are the same as
in Fig.\protect~\ref{fig:small}.}
\label{fig:spectra}
\end{figure*}

The photoabsorption spectra of these clusters are represented in
Fig.~\ref{fig:spectra} for increasing temperatures in the range $0\leq T\leq
60$~K for all sizes except CaAr$_{37}$, for which $0\leq T\leq 40$~K. The
classical spectra have also been plotted as dashed lines for $T>0$. Except for
CaAr$_{37}$, the classical spectrum looks very much like its quantum
counterpart with slightly shifted but noticeably narrower excitation peaks.
The four clusters display different absorption spectra,
characteristic of finite-size effects. In CaAr$_{13}$, the spectrum gradually
evolves from a two-peak structure (at about 23700~cm$^{-1}$ and
24100~cm$^{-1}$) to another two-peak structure (at about
23650 and 23800 cm$^{-1}$). The peak ending at
23800~cm$^{-1}$ appears first at 30~K, and originates from
isomers where the Ca atom lies in the main icosahedral shell. No real signature
of melting at 40~K is seen in Fig.~\ref{fig:spectra}(a), however the present
spectra can still be interpreted in terms of isomerizations within the cluster.

In the
case of CaAr$_{30}$, the low temperature isomerization below 5~K has a strong
effect on the absorption peaks, which are red shifted by about 200 wavenumbers.
The spectra of this cluster also evolves very continuously, and the two main
peaks at 23550 and 24000~cm$^{-1}$ undergo a slight red shift and thermal
broadening. The weak melting phase change is reflected on these spectra, and
the most prominent feature remains the sharp isomerization below 5~K.

\begin{figure*}[htb]
\vbox to 7cm{
\includegraphics{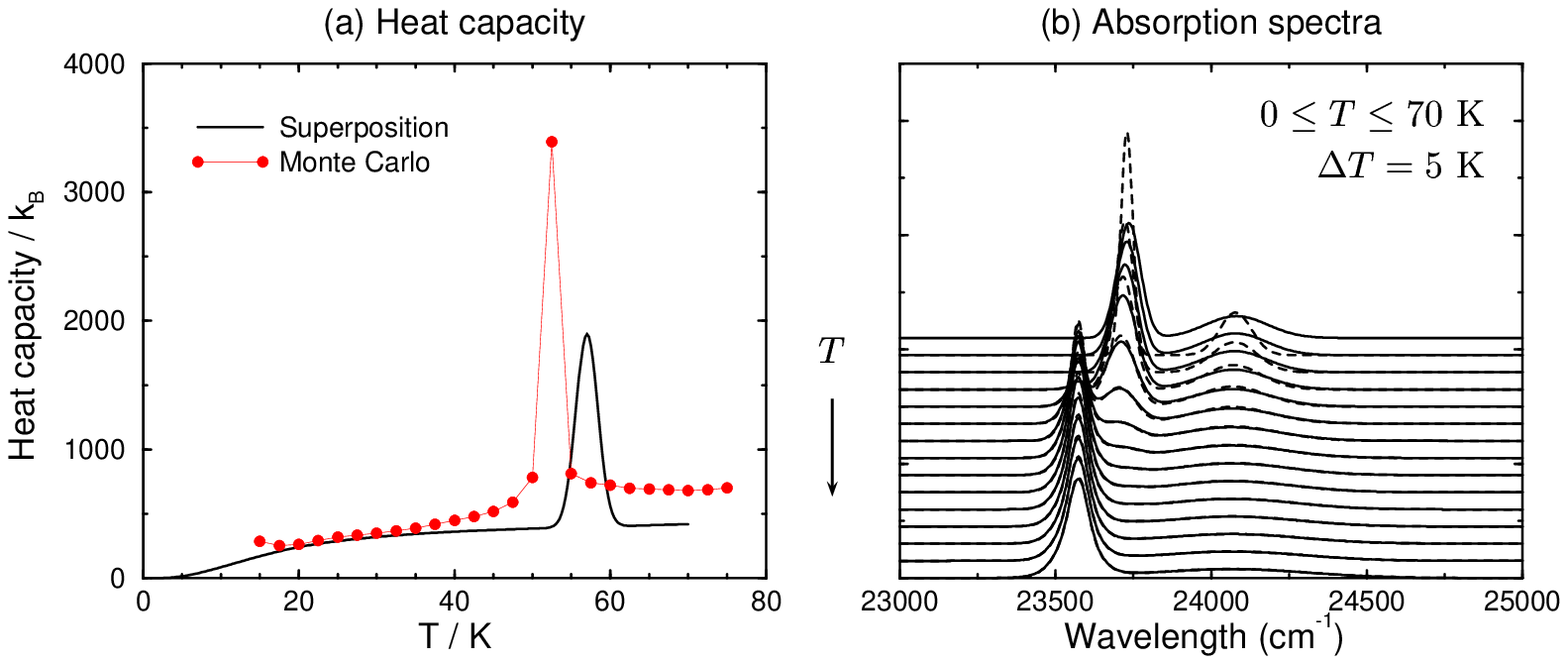}
\vfill}
\caption{(a) Heat capacity and (b) absorption spectra of CaAr$_{146}$.
The notations are the same as in Fig.\protect~\ref{fig:small}.}
\label{fig:caar146}
\end{figure*}

Near $T=10$~K, the multiple peak structure of the spectrum in CaAr$_{37}$ also
reveals that several isomers are coexisting with distinct spectroscopic
signatures. 
The solid-solid transition seen in the heat capacity of CaAr$_{37}$ at low
temperature is reflected on the absorption spectra in
Fig.~\ref{fig:spectra}(c). The three-peak $T=0$~K spectrum is peculiar to the
decahedral ground state. At $T\sim 10$~K, a significant blue shift
occurs in the middle peak, providing a signature of the icosahedral isomers.
The quantitative difference in the solid-solid transition temperature induced
by quantum effect is also seen in the absorption spectra between 5 and 15~K.
Near 15~K, the spectrum can be clearly separated into more than three peaks,
and suggests that distinct icosahedral isomers are present. At around 20~K,
the various peaks merge into a double-peak structure, where the peak at
23800~cm$^{-1}$ dominates the one at 24750~cm$^{-1}$. Here we find a
clear thermodynamical influence on the spectroscopic properties. Hence,
in this cluster,
the absorption spectrum contains notable informations about isomerization and
phase changes, including some rather fine details on the characteristic
temperatures.

The equilibrium geometry of CaAr$_{54}$ is a double layer Mackay icosahedron
where the Ca atom is located on a vertex site, thus keeping a relatively high
symmetry (C$_{5v}$). As in Ar$_{55}$, the heat capacity near peak 40~K
indicates volume melting,\cite{labastie,wales93} while preliminary surface
melting has a very small thermal signature. In Fig.~\ref{fig:spectra}(d) we
observe that the two absorption peaks undergo a significant change near 30~K,
and this corresponds to coexisting isomers with the Ca atom located on
different substitutional sites of the external icosahedral shell. In this case
the photoexcitation spectrum is sensitive to surface melting. At higher
temperatures $T>40$~K, many isomers are found in which the metal atom is more
fully solvated by argon. The spectra show another qualitative change, as the
peak near 23600~cm$^{-1}$ disappears and is replaced by a broad bump near
24300~cm$^{-1}$.

\subsection{\boldmath CaAr$_{146}$\unboldmath}

It is interesting to compare the situation in CaAr$_{54}$ with the one in the
larger Mackay icosahedron, CaAr$_{146}$. As in CaAr$_{54}$, the ground state
equilibrium geometry has C$_{5v}$ symmetry, with calcium located in a vertex
site of the external icosahedral layer. This cluster has a huge number of
isomers, and many high lying minima are missed during the quenching process.
In the vast majority of
structures in our sample the calcium atom remains in the outer argon layers.
This is not surprising, since, and as was previously mentioned,
the calcium-argon bond is weaker by about 70\%
than the argon-argon bond.\cite{epjd}
Despite the relatively small size of our sample with respect to the number of
existing structures, the heat capacities in Fig.~\ref{fig:caar146}(a) exhibit
a reasonably
good agreement between simulation and the superposition method, the high peak
near 55~K being the signature of the solidlike-liquidlike phase change. Melting
is essentially caused by the motion of argon atoms, and is relatively
unsensitive to the location of calcium. This explains the shape and the
evolution of the absorption spectrum in Fig.~\ref{fig:caar146}(b). The two
peaks at 23720 and 24150~cm$^{-1}$ become broader and vanish between 25~K
and 30~K, to be replaced by a major peak at 23600~cm$^{-1}$. This transition is
correlated with fluctuations in the location of the calcium atom over the
cluster surface. On the other hand, melting at 55~K does not lead to any
visible change in the absorption spectrum.

\subsection{Discussion}

The above results may not be immediately amenable to experimental comparison,
because of the problem of size selection. Our simulation results show that
complex isomerizations involving strong changes in the local environment of
calcium could be, in principle, detected in experiments on size-selected
clusters initially thermalized. However a typical sample of clusters in
molecular beams is made of a broad size distribution, where the width and
the center have values close to each other. The common feature
of these clusters is the surface location of the
calcium atom, and we may expect the absorption spectrum to be a possible probe
of surface melting in these clusters, provided that the surface melting
temperature does not significantly depend on cluster size. Even though calcium
in bulk argon has a very distinct, three-peak spectroscopic signature,%
\cite{epjd}
it may be hard to detect volume melting in which calcium would be fully
solvated. In fact, such van der Waals clusters evaporate atoms at temperature
close to the melting point. From a statistical point of view, the weakly
bonded calcium atom is likely to evaporate first. In addition the situations
where calcium is fully solvated are less probable. Therefore the true volume
melting transition occuring with full solvation of calcium may hardly be
observed.

Provided that one could size-select these clusters, we have shown in the
present work that spectroscopy can possibly probe their phase
changes. More generally, intricate finite size effects in the thermodynamical
behavior could be expected to be seen,
as in CaAr$_{13}$, CaAr$_{30}$ or CaAr$_{37}$. In these special clusters,
coexisting isomers and multiple phase-likes
induce complex variations of the absorption properties with
increasing temperature.

Even with experimental setups involving selected sizes, another problem may
come into play to hinder one from getting the complex spectroscopic
information. All the
above results are supposed to be at thermal equilibrium, without assuming how
this equilibrium state had been reaching. In some cases, the relaxation
time required for achieving equilibration may be very long because of multiple
funnels in the energy landscape and kinetic trapping.
This has been observed in the Ar$_{38}$ cluster by Miller and coworkers,%
\cite{ar38miller} where typical timescales
of 10~s have been estimated for the interfunnel transitions at moderate
temperatures. Relaxation to equilibrium and time-dependent absorption spectra
will be the subject of the following article.\cite{papdjw}

\section{Conclusion}
\label{sec:ccl}

In this paper, we  have investigated some methodological and practical
aspects  of finite temperature  spectroscopy in  chromophore-doped van
der  Waals  clusters  at  thermal  equilibrium.   Using  the  harmonic
superposition   approach   improved   with   anharmonic   perturbation
corrections  to  the  partition  function, the  basic  thermodynamical
observables were calculated from samples of local minima on the energy
landscape. The photoabsorption spectrum  was obtained using the recent
Gaussian  theory  by  Wadi   and  Pollak  \cite{wp}  in  the  harmonic
approximation. These  theories were  tested and validated  on selected
CaAr$_n$ clusters, in  both the classical and quantum  regimes, at low
or moderate  temperatures. Although anharmonic  corrections could also
be incorporated to the various  quantities (line shifts and widths) in
this theory, we found them  unnecessary in the present work because of
the other underlying approximations.

We have  studied the possible spectroscopic signatures  of the various
isomerizations   and   phase   changes   in   CaAr$_6$,   CaAr$_{10}$,
CaAr$_{13}$, CaAr$_{30}$,  CaAr$_{37}$, CaAr$_{54}$, and CaAr$_{146}$.
The two smaller clusters show relatively simple isomerizations leading
to a similar  spectroscopic signature with a clear  three- to one-peak
pattern.   Both  CaAr$_{30}$   and  CaAr$_{37}$  display  strong  (non
monotonic)  finite-size  effects   and  solid-solid  transitions.   In
CaAr$_{13}$, melting  involves the calcium  atom taking place of
one argon atom  in the icosahedral shell. In  contrast with most other
sizes, it is associated with  the appearance of a new absorption peak.
In CaAr$_{54}$ and CaAr$_{146}$, preliminary surface melting occurs at
nearly  30~K, and  is reflected  in the  absorption spectrum.   In all
these  phenomena,  the local  environment  of  calcium changes  rather
abruptly,  resulting  usually  in  clear  shifts  in  some  absorption
peaks. Volume melting  towards a calcium atom fully  solvated by argon
has a  potentially significant spectral signature,  however calcium is
less  bound  to  argon  than  argon  itself, hence  it  is  likely  to
dissociate before  volume melting occurs. One should  mention that the
main heat capacity  peak indicates melting of the  argon cluster host,
especially for large sizes.  While  the host cluster is liquidlike the
calcium  is still  free, to  some extent,  to glide  over  the cluster
surface.   Being in this  state, the  rearrangements inside  the argon
cluster  have  little  impact   on  the  global  spectroscopy  of  the
chromophore,  and the system  can be  expected to  behave as  a pseudo
diatomics Ca--Ar$_n$. This point will be examined further in the third
paper.\cite{papjmm}

The techniques used here are valid in both the quantum and classical regimes,
which refers to the limit $\hbar\to 0$.
They could be applied to other chromophore-doped van der Waals clusters,
provided that potential energy surfaces are built first. In particular, neon
clusters could be profitably investigated using the superposition method and
the Gaussian theory of absorption. For helium, the harmonic approximation may
break down, and even anharmonic corrections should be unsufficient. In this
case, quantum Monte Carlo should be coupled with on-the-fly histogram
accumulation to compute the spectrum.
More generally, the present methods are relevant to any statistical driven
absorption spectrum in heterogeneous or homogeneous polyatomic molecules.
For instance, the thermal effects on the absorption spectrum in small sodium
clusters \cite{moseler,schmidt} could be investigated using appropriate
models.\cite{ddtb,naab} As concerns the approximations within the Gaussian
theory, the most questionable point is the short-time expansion in the
autocorrelation function. This approximation could be overcome using the
formal results of Yan and Mukamel.\cite{yan}
Unfortunately, this method involves a heavy numerical effort in the
context of multiple isomers, and it still assumes harmonic forces. Finally,
the superposition approximation provides a framework to study time-dependent
properties over long time scales not available to molecular dynamics,
by solving appropriate master equations.\cite{med} The application of this
technique to photoabsorption spectra will be made in the next paper of this
series.\cite{papdjw}

\section*{Acknowledgments}

We thank CALMIP for a generous allocation of computer resources. FC also
acknowledges interesting discussions with Dr. J. Vigu\'e.

\appendix
\section*{Appendix: Perturbation expansion of the partition function}
\label{app:perturbpart}

The ground state Hamiltonian is written in cartesian coordinates ${\bf P},
{\bf R}$ as $H({\bf P},{\bf R}) = H_a({\bf P},{\bf R}) + \delta V({\bf R})$,
where $H_a$ is the harmonic term and $\delta V$ the extra, anharmonic
contribution. The Taylor expansion of $\delta V$ near the minimum ${\bf
R}_0$ contains products of at least 3 coordinates. The partition
function $Z(\beta)={\rm Tr}\, \exp(-\beta H)$ is now expanded in a
semiclassical fashion, in order to yield at least the same terms as the
corresponding
classical expansion. This approximation is actually correct from the quantum
point of view as long as we keep it to first order. However, the rigorous
perturbation treatment, which should be performed with path integrals, will
not allow us to obtain the simple corrections that can be compared with the
classical expressions of Ref.~\onlinecite{jcpanhar}. Thus we write in a
phenomenological way
\begin{equation}
Z(\beta) \approx {\rm Tr}\, \left[ e^{-\beta H_a} \prod_{n=0}^2
\frac{(-\beta)^n}{n!}(\delta V)^n\right],
\label{eq:ap1}
\end{equation}
and we keep only terms containing coordinates up to the power 12, which
contribute to $T^2$ corrections in the low-temperature classical partition
function.\cite{jcpanhar} As in the classical case, all odd products have a
zero contribution, but it is no longer possible to sort the various nonzero
terms according to their power in $\beta$. It appears more natural to sort them
according to the even power in coordinates. Following our previous notations%
\cite{jcpanhar} we write the anharmonic term $\delta V$ as a function of the
normal mode coordinates ${\bf Q}$ as
\begin{eqnarray}
\delta V({\bf Q}) &=& \sum_{k\geq 3} \frac{1}{k!} \sum_{i_1\cdots i_k}\left.
\frac{\partial^k V}{\partial Q_{i_1}\cdots \partial Q_{i_k}}\right|_{{\bf Q}=0}
Q_{i_1}\times \cdots \times Q_{i_k} \nonumber \\
&=& \sum_{k\geq 3}\frac{1}{k!} \sum_{i_1\cdots i_k} \Gamma_{i_1\cdots
i_k}^{(k)} Q_{i_1}\times \cdots \times Q_{i_k}.
\label{eq:app6}
\end{eqnarray}
The corrective terms can be expressed as averages on the harmonic ground state:
\begin{equation}
Z(\beta)\approx Z_0(\beta)[1+\gamma_4+\gamma_6+\gamma_8+\gamma_{10}+
\gamma_{12}],
\label{eq:app7}
\end{equation}
where $\gamma_i$ labels the correction due to power $i$ in coordinates. These
terms are explicitely given by
\begin{eqnarray}
\gamma_4 &=&- \frac{\beta}{24}\sum_{i_1\cdots i_4} \Gamma_{i_1\cdots i_4}^{(4)}
\langle Q_{i_1}\times \cdots \times Q_{i_4}\rangle;
\label{eq:gamma4} \\
\gamma_6 &=& \sum_{i_1\cdots i_6} \left[ \frac{-\beta}{720}
\Gamma_{i_1\cdots i_6}^{(6)} + \frac{\beta^2}{72}\Gamma_{i_1i_2i_3}^{(3)}
\Gamma_{i_4i_5i_6}^{(3)}
\right] \langle Q_{i_1}\times \cdots \times Q_{i_6}\rangle \label{eq:gamma6};\\
\gamma_8 &=& \sum_{i_1\cdots i_8} \left[ \frac{\beta^2}{1152}
\Gamma_{i_1\cdots i_4}^{(4)}\Gamma_{i_5\cdots i_8}^{(4)}
+ \frac{\beta^2}{720}\Gamma_{i_1i_2i_3}^{(3)}
\Gamma_{i_4\cdots i_8}^{(5)}
\right] \langle Q_{i_1}\times \cdots \times Q_{i_8}\rangle \label{eq:gamma8};\\
\gamma_{10}&=&\sum_{i_1\cdots i_{10}} \left[ \frac{\beta^2}{28800}
\Gamma_{i_1\cdots i_5}^{(5)}\Gamma_{i_6\cdots i_{10}}^{(5)}
+ \frac{\beta^2}{17280}\Gamma_{i_1\cdots i_4}^{(4)}
\Gamma_{i_5\cdots i_{10}}^{(6)} \right.\nonumber \\
&& \left. - \frac{\beta^3}{1728}\Gamma_{i_1i_2i_3}^{(3)}
\Gamma_{i_4i_5i_6}^{(3)}\Gamma_{i_7\cdots i_{10}}^{(4)}
\right] \langle Q_{i_1}\times \cdots \times Q_{i_{10}}\rangle;
\label{eq:gamma10}\\
\gamma_{12}&=&\sum_{i_1\cdots i_{12}} \left[ \frac{\beta^2}{1036800}
\Gamma_{i_1\cdots i_6}^{(6)}\Gamma_{i_7\cdots i_{12}}^{(6)}
- \frac{\beta^3}{51840}\Gamma_{i_1i_2i_3}^{(3)}\Gamma_{i_4i_5i_6}^{(3)}
\Gamma_{i_7\cdots i_{12}}^{(6)} \right. \nonumber \\
&&- \frac{\beta^3}{17280}\Gamma_{i_1i_2i_3}^{(3)}
\Gamma_{i_4\cdots i_7}^{(4)}\Gamma_{i_8\cdots i_{12}}^{(5)}
- \frac{\beta^3}{82944}\Gamma_{i_1\cdots i_4}^{(4)}
\Gamma_{i_5\cdots i_8}^{(4)}\Gamma_{i_9\cdots i_{12}}^{(4)} \nonumber \\
&&
\left. + \frac{\beta^4}{31104}\Gamma_{i_1i_2i_3}^{(3)}\Gamma_{i_4i_5i_6}^{(3)}
\Gamma_{i_7i_8i_9}^{(3)}\Gamma_{i_{10}i_{11}i_{12}}^{(3)}
\right] \langle Q_{i_1}\times \cdots \times Q_{i_{10}}\rangle.
\label{eq:gamma12}
\end{eqnarray}
The above expressions can be considerably simplified by factorizing the
average products $\langle Q_i\times \cdots \times Q_j\rangle$. This is made
possible by Wick's theorem, which holds for general bosonic ground states:
\begin{equation}
\langle Q_i^{2n}\rangle = \frac{1}{2^n}\frac{(2n)!}{n!}\langle Q_i^2\rangle^n
= (2n-1)!! \langle Q_i^2\rangle^n,
\label{eq:app9}
\end{equation}
with
\begin{equation}
\langle Q_i^2\rangle = \frac{\hbar}{\omega_i} \left( \frac{1}{2} + \frac{1}
{e^{\beta \hbar \omega_i}-1}\right).
\label{eq:app9bis}
\end{equation}
It is then more convenient to write the $\gamma_i$'s as diagrams, in which each
$k-$vertex represents one of the $\Gamma_{i_1\cdots i_k}^{(k)}$ term. For
instance we have
\begin{equation}
\gamma_4 = -\frac{\beta}{8}\diagg{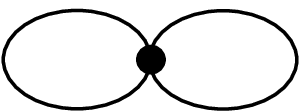}{1.5}{0.8}
\label{eq:app10a}
\end{equation}
with
\begin{equation}
\diagg{D4_1}{1.5}{0.8}=\sum_{ij}\Gamma_{iijj}^{(4)}\langle Q_i^2\rangle \langle
Q_j^2\rangle.
\label{eq:app10b}
\end{equation}
The two diagrams giving $\gamma_6$ are
\begin{equation}
\gamma_6=-\frac{\beta}{48}\diagg{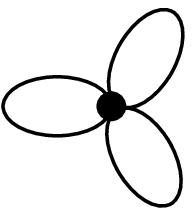}{1}{2.2}
+\frac{\beta^2}{8}\diagg{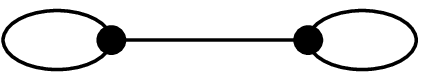}{2}{0.2}
+ \frac{\beta^2}{12}\diagg{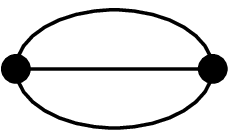}{1.2}{1}.
\label{eq:app11}
\end{equation}
The other perturbative terms $\gamma_8$, $\gamma_{10}$, and $\gamma_{12}$
contain a larger number of diagrams, including non connected ones. The fewest
diagrams are obtained in $\ln Z$, which still has 46 different contributions
corresponding to Eq.~(\ref{eq:app7}):
\begin{widetext}
\begin{eqnarray}
\ln Z = \ln Z_0 && -
\frac{\beta}{8} \diagg{D4_1}{1.5}{0.8} - \frac{\beta}{48} \diagg{D6_1}{1}{2.2}
+ \frac{\beta^2}{8} \diagg{D33_1}{2}{0.2} +
\frac{\beta^2}{12}\diagg{D33_2}{1.2}{1}
\nonumber \\
&& + \frac{\beta^2}{48} \diag{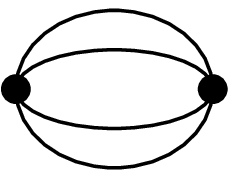}{2.5}+
\frac{\beta^2}{16}\diagp{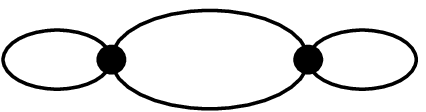}{2.5} +
\frac{\beta^2}{16} \diag{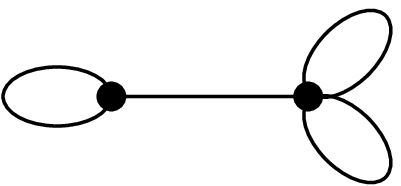}{2.5}+\frac{\beta^2}{12}\diagp{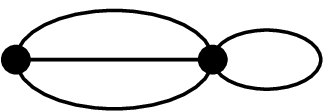}{2.5}
\nonumber \\
&& + \frac{\beta^2}{128}\diagq{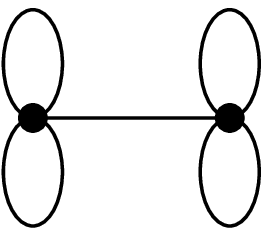}{2.5}+\frac{\beta^2}{48}\diagp{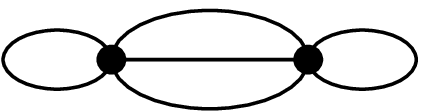}{2.5}
+\frac{\beta^2}{240} \diag{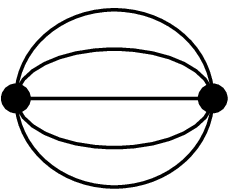}{2.5} + \frac{\beta^2}{32}\diag{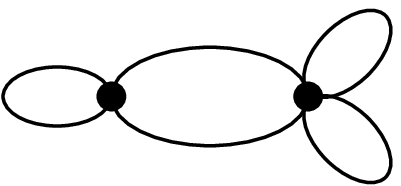}{2.5}
\nonumber \\
&& + \frac{\beta^2}{48} \diag{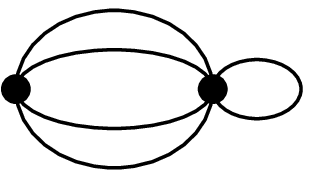}{2.5} - \frac{\beta^3}{16}
\diagq{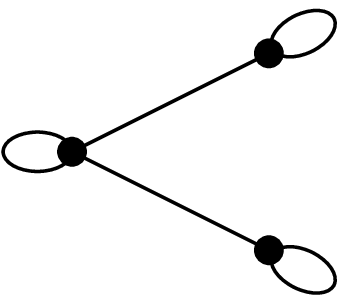}{2.5} - \frac{\beta^3}{8} \diagq{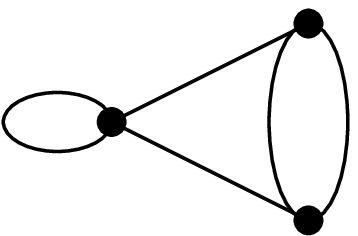}{2.5}-
\frac{\beta^3}{8}\diagq{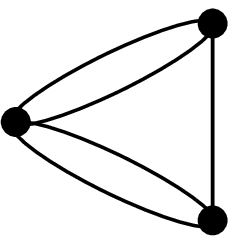}{2.5} \nonumber \\
&& - \frac{\beta^3}{8} \diagq{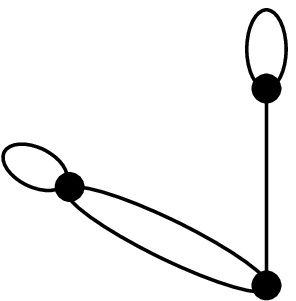}{2.5} - \frac{\beta^3}{12}
\diagq{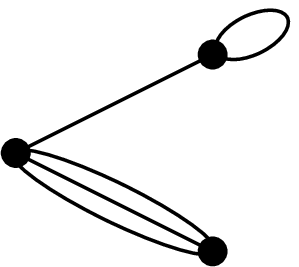}{2.5} + \frac{\beta^2}{256} \diagq{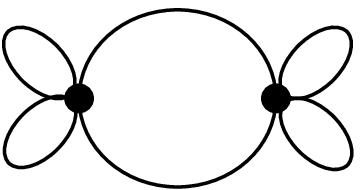}{2.5} +
\frac{\beta^2}{192}\diag{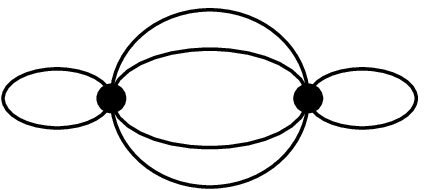}{2.5} \nonumber \\
&& + \frac{\beta^2}{1440} \diagq{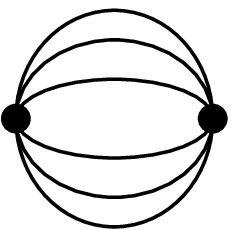}{2.5} - \frac{\beta^3}{64}
\diagq{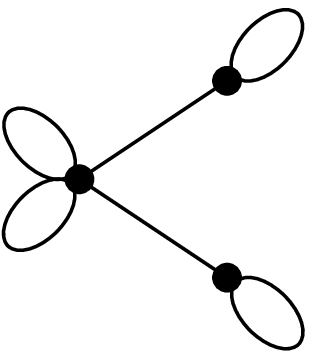}{2.5} - \frac{\beta^3}{24} \diagq{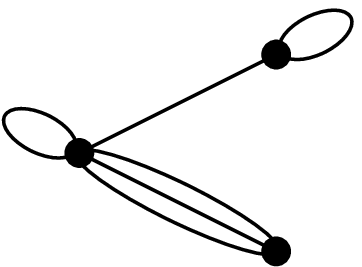}{2.5} -
\frac{\beta^3}{32}\diag{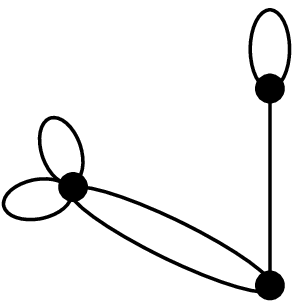}{2.5} \nonumber \\
&& - \frac{\beta^3}{32} \diagq{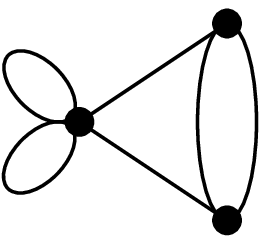}{2.5} - \frac{\beta^3}{16}
\diagq{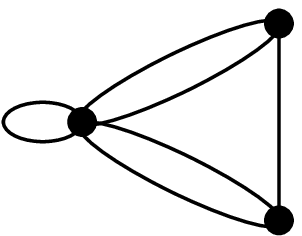}{2.5} - \frac{\beta^3}{72} \diagq{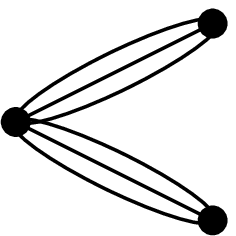}{2.5} -
\frac{\beta^3}{32}\diagq{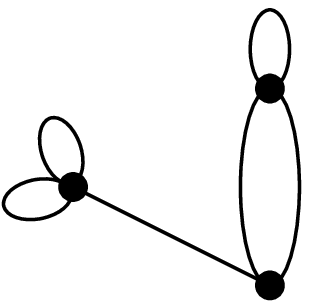}{2.5} \nonumber \\
&& - \frac{\beta^3}{8} \diagq{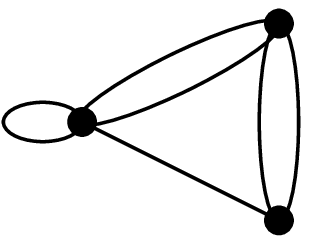}{2.5} - \frac{\beta^3}{48}
\diagq{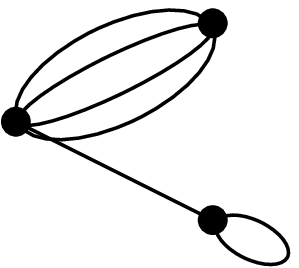}{2.5} - \frac{\beta^3}{32} \diagq{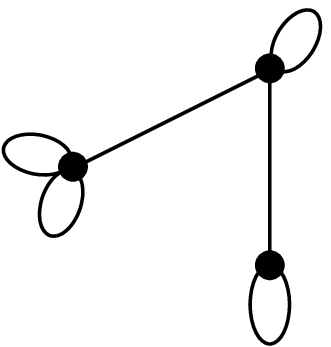}{2.5} -
\frac{\beta^3}{8}\diagq{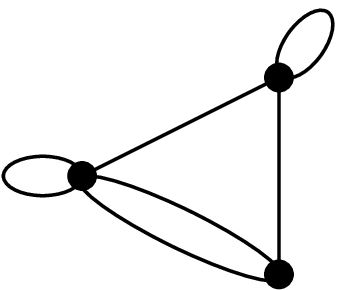}{2.5} \nonumber \\
&& - \frac{\beta^3}{12} \diagq{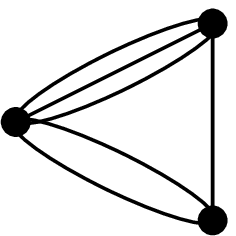}{2.5} - \frac{\beta^3}{24}
\diagq{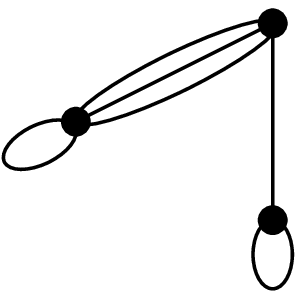}{2.5}- \frac{\beta^3}{24} \diagq{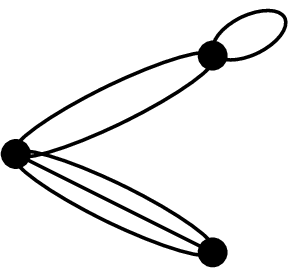}{2.5} -
\frac{\beta^3}{16}\diagq{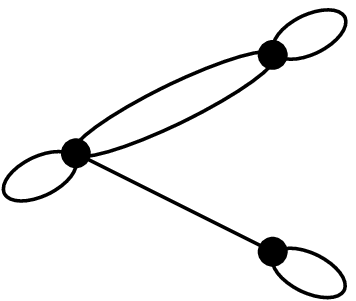}{2.5} \nonumber \\
&& - \frac{\beta^3}{48} \diagq{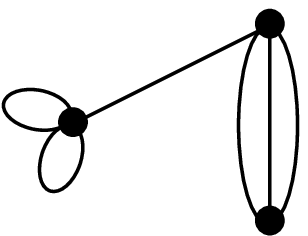}{2.5} -
\frac{\beta^3}{48}\diagq{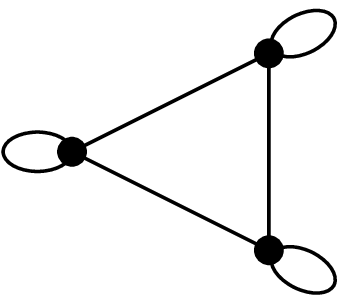}{2.5}- \frac{\beta^3}{24} \diagq{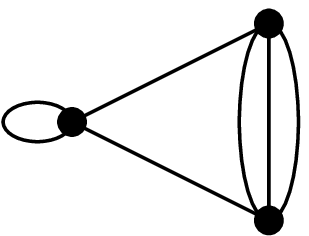}{2.5}
- \frac{\beta^3}{32}\diagq{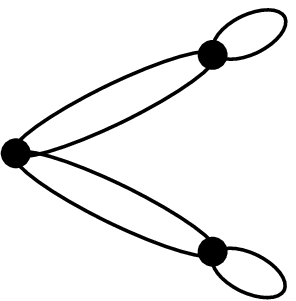}{2.5} \nonumber \\
&& - \frac{\beta^3}{48} \diagq{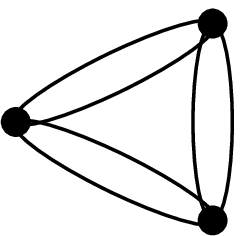}{2.5} + \frac{\beta^4}{16}
\diagq{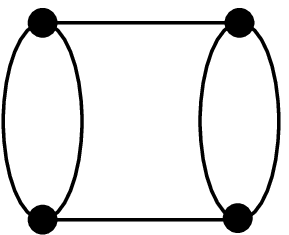}{2.5}+ \frac{\beta^4}{16} \diagq{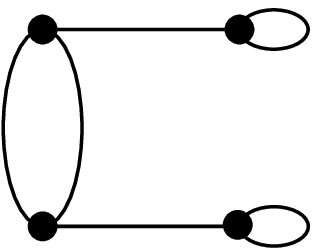}{2.5} +
\frac{\beta^4}{48}\diagq{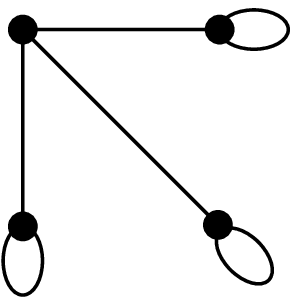}{2.5} \nonumber \\
&& + \frac{\beta^4}{8} \diagq{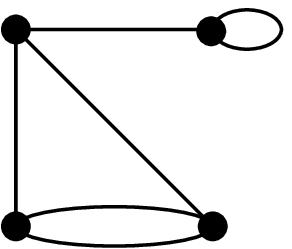}{2.5} +
\frac{\beta^4}{24}\diagq{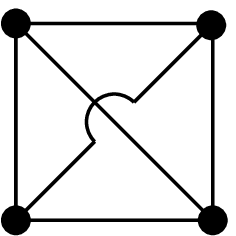}{2.5} \label{eq:alp2}.
\end{eqnarray}
\end{widetext}
It is important to notice that this perturbative expansion includes all 18
terms up to second-order in the classical expansion,\cite{jcpanhar} and that
28 remaining terms would contribute classically to 3rd and higher orders in
$T^n$. These terms correspond to the $\Gamma^{(5)}\times \Gamma^{(5)}$,
$\Gamma^{(4)}\times \Gamma^{(6)}$,
$\Gamma^{(3)}\times \Gamma^{(4)}\times \Gamma^{(5)}$, and
$\Gamma^{(4)}\times \Gamma^{(4)}\times \Gamma^{(4)}$ products, but cannot be
neglected a priori with respect to the other corrections in the quantum regime.

In the superposition method framework, these perturbative corrections act on
the individual partition functions, where they are expected to give a more
accurate estimate of the relative weights of the isomers as a function of
temperature, as well as more accurate thermodynamical observables.

\end{document}